\documentclass[11pt]{article}

\usepackage[utf8]{inputenc}
\usepackage[T1]{fontenc}
\usepackage{lmodern}
\usepackage{csquotes}
\usepackage[UKenglish]{babel}
\usepackage[a4paper, margin=1in]{geometry}
\usepackage{setspace}
\usepackage{graphicx}
\usepackage[section]{placeins}

\usepackage{booktabs}
\usepackage{makecell}
\usepackage{array}
\usepackage{tabularx}
\usepackage{longtable}
\usepackage{threeparttable}
\newcolumntype{Y}{>{\centering\arraybackslash}X}
\providecommand{\sym}[1]{\ifmmode^{#1}\else\(^{#1}\)\fi}

\newcolumntype{L}{>{\raggedright\arraybackslash}X}
\newcolumntype{P}[1]{>{\raggedright\arraybackslash}p{#1}}

\usepackage{amsmath,amssymb}

\usepackage{appendix}
\usepackage{chngcntr}

\usepackage[style=apa,sorting=nyt]{biblatex}
\usepackage[protrusion=true,expansion=true]{microtype}
\usepackage{newpxtext,newpxmath}
\usepackage{hyperref}
\usepackage[nameinlink,capitalize,noabbrev]{cleveref}
\hypersetup{colorlinks=true, linkcolor=blue, citecolor=blue, urlcolor=blue}

\usepackage{authblk}

\usepackage{listings}
\usepackage{xcolor}
\definecolor{promptbg}{RGB}{248,248,248}
\lstdefinestyle{prompt}{
   basicstyle=\ttfamily\fontsize{6.2pt}{7.4pt}\selectfont,
   backgroundcolor=\color{promptbg},
   frame=single,
   rulecolor=\color{black!20},
   breaklines=true,
   breakautoindent=false,
   breakindent=0pt,
   columns=fullflexible,
   keepspaces=true,
   tabsize=2,
   xleftmargin=4pt,
   xrightmargin=4pt,
   aboveskip=6pt,
   belowskip=6pt,
}

\newcommand{\papertitle}{\textbf{Nine Raters, One Index: Carrying LLM Disagreement into Labour-Market Estimates}}

\newcommand{\paperabstract}{%
When a large language model supplies research annotations, the choice of model becomes an analytic degree of freedom. We ask how much that choice matters by putting 1,100 task--occupation cells to nine models from three vendors and weighting their ratings with worker-reported task importance in the British Skills and Employment Surveys. Rankings are substantially more stable than levels: pairwise rank correlations range from 0.74 to 0.92, while the share of British jobs scoring above 0.5 ranges from under 0.1\% to 38\%. We therefore treat the rater as one dimension of a multiverse, reporting each rater-specific estimate and its range alongside a Rubin-style summary that incorporates their observed dispersion. The consequences depend on the downstream design. A post-2022 pay gradient of $-0.020$ log points per interquartile range is negative under all nine raters, although it continues a pre-existing trend. In online vacancies, all nine raters initially produce
negative post-2022 gradients; after teleworkability is given its own period path, seven produce significant positive coefficients, two produce small negative coefficients, and the pooled estimate is indistinguishable from zero. Neither application identifies an effect of generative AI. Validation against worker-reported AI use, competing exposure measures, assistant traffic and an independent task survey supports incremental predictive content and portability, but any bias shared across raters remains unidentified. The paper's contribution is a procedure for exposing and carrying observed LLM-rater disagreement into empirical estimates.
}

\newcommand{\paperkeywords}{large language models, LLM-as-rater, measurement uncertainty, multiverse analysis, task-based approach, labour market exposure}

\title{\papertitle}

\author[1]{Golo Henseke\thanks{Corresponding author: Golo Henseke, University College London, 20 Bedford Way, London WC1H 0AL, UK. Email: \texttt{g.henseke@ucl.ac.uk}.}}
\author[2]{Rhys Davies}
\author[2]{Alan Felstead}
\author[3]{Duncan Gallie}
\author[1]{Francis Green}
\author[4]{Ying Zhou}

\affil[1]{University College London, 20 Bedford Way, London WC1H 0AL, UK}
\affil[2]{Cardiff University, Maindy Road, Cardiff CF24 4HQ, UK}
\affil[3]{Nuffield College, University of Oxford, New Road, Oxford OX1 1NF, UK}
\affil[4]{Surrey Business School, University of Surrey, Alexander Fleming Road, Guildford GU2 7XH, UK}

\date{}
\hypersetup{
  pdftitle={Nine Raters, One Index: Carrying LLM Disagreement into Labour-Market Estimates},
  pdfauthor={Golo Henseke, Rhys Davies, Alan Felstead, Duncan Gallie, Francis Green, Ying Zhou}
}

\newcommand{\TabOneCaption}{Task Categories by Broad Domain}
\newcommand{\TabTwoCaption}{Task Importance Scores}
\newcommand{\TabFiveCaption}{Where Occupations Sit in the Exposure Ordering}
\newcommand{\TabSevenCaption}{Average 'Coverage' Share by Major SOC Occupation (weighted)}
\newcommand{\TabNineCaption}{Average Marginal Effects of Exposure Indices on AI Adoption}

\newcommand{\FigOneCaption}{GAISI Classification Pipeline}
\newcommand{\FigFourCaption}{Predicted Probability of AI Use by GAISI Quintile.}
\newcommand{\FigFiveCaption}{Change in Mean Exposure Rank, 2017--2023/24.}
\newcommand{\FigSixCaption}{Online Job Adverts and Occupational Exposure to Generative AI, 2017--2026.}

\begin{document}

\setcounter{page}{1}

\maketitle

\begin{abstract}
\setstretch{1.15}
\paperabstract
\end{abstract}


\bigskip

\section{Introduction}
The share of British jobs that a large language model rates as highly exposed to generative AI ranges from under a tenth of one per cent to 38 per cent. The only difference between those two numbers is which model is asked. The assessment rubric, the tasks, their importance weights, and the occupational context are identical; the raters are nine frontier models from multiple generations, and yet, their answers to ``how much of this job could an AI assistant meaningfully augment'' are that far apart.

Exposure ratings of this kind underpin a growing empirical literature, and the question they address is not just academic. AI adoption has been fast: by early 2026, 36\% of UK workers and 43\% of US workers reported using generative AI at work \parencite{Bick2026MindU.S.}, and already in mid-2024 some 7.4 million British workers, a quarter of the workforce, reported using AI-enabled software of some kind \parencite{Henseke2025WhatAdoption}. Exposure measures are intended to rank work by its susceptibility to a
technology. Whether that susceptibility predicts adoption or subsequent labour-market change is a separate empirical question \parencite{Acemoglu2025TheAI,Korinek2024TheAI}. Measures of technology exposure typically rate job tasks for their susceptibility to a specific technology, then aggregate the ratings additively to occupations \parencite{Felten2021OccupationalUses,Webb2020TheMarket,Brynjolfsson2018WhatEconomy}. For LLMs, the assessment of \textcite{Eloundou2024GPTsLLMs} (hereafter EMMR) has become the standard reference. Its LLM-rated version has subsequently been used in a range of empirical
applications \parencite[e.g.][]{eisfeldt2023GenAI,KleinTeeselink2025GenerativeKingdom, Brynjolfsson2025CanariesTo}. However, when the rating instrument is itself a model, the choice of model becomes a researcher degree of freedom \parencite{Breznau2022manyresearcher,stureborg2024llminconsistent} that published estimates rarely report the consequences of. This paper asks what those consequences are. We show they are not intrinsic to the index: they depend on the design the index is used in, from negligible where other sources of variation restrict the estimate to dominating where sampling error is nearly exhausted.

We rate each of 1,100 task--occupation cells with a panel of nine LLM raters --- Gemini 1.5, 2.5 and 3.1 Pro, GPT-4o, 5.1, 5.4, 5.5 and 5.6-Luna, and GLM 5.2 --- and weight the ratings by worker-reported task importance in the British Skills and Employment Surveys, yielding a worker-level index we call GAISI. All substantive estimates are then produced nine times over, once per rater, and reported as a range as well as a pooled figure in which disagreement between raters enters the standard error. Reporting the whole set is a multiverse analysis \parencite{Steegen2016IncreasingAnalysis,Simonsohn2020SpecificationDescriptions} with the rater as the varied degree of freedom; the pooling computes a Rubin-combined variance to provide a pragmatic summary of statistical uncertainty \parencite{Rubin1987MultipleSurveys}.

Three main findings stand out. First, rankings are substantially more stable across raters than levels. Pairwise rank correlations between the nine range from 0.74 to 0.92, and averaging over the panel raises reliability from 0.75 for a single rater to 0.96. By contrast, the share of British jobs scoring above 0.5 is the 0.1-to-38 per cent spread above. Exposure therefore enters every substantive analysis in this study as a percentile rank. 

Second, the facets of the rating process differ in how much they matter. Re-running one model on one prompt with a different random seed leaves the ranking almost intact. Switching sampling off does not materially change the ranking. Other changes are more consequential. Moving between model versions inside one vendor's lineage disrupts ratings as much as moving between vendors. A panel assembled to span vendors but not generations can understate the problem.

Third, the labour-market applications document post-2022 gradients but do not identify an effect of generative AI. In the Labour Force Survey, the pay gap between more and less exposed occupations narrows by 0.020 log points per interquartile range of exposure, but the decline began before ChatGPT. New postings also move against more exposed occupations after 2022, but the series was already trending, and the negative association disappears when teleworkability is given its own period path. The two applications differ sharply in their sensitivity to the rater. The between-rater share of the Rubin-combined variance accounts for about a twentieth of the uncertainty in the pay estimate, which is negative under all nine raters, but 92 per cent in the baseline vacancy estimate and 98 per cent after controlling for teleworkability. Under the latter specification, seven raters produce significant positive post-2022 coefficients, two produce small negative coefficients, and the pooled estimate is not distinguishable from zero. The descriptive pay gradient is robust across the rater ensemble; the telework-adjusted vacancy estimate is not.

Alongside the method, we build the index and assess whether it carries predictive information beyond existing exposure measures and the SES setting. Exposure concentrates in cognitive, high-skill work --- high-skill occupation groups sit 37 percentile points above the lower-skilled ones --- reversing the gradient of earlier automation waves, and it rose 4.5 percentile points between 2017 and 2023/24 almost entirely through changes in which occupations people hold rather than in what those jobs involve. We compare the index with workers' reported AI use, established exposure measures, observed assistant traffic, and an independent task instrument. Together these checks assess incremental predictive content and portability, but any bias shared across raters remains unidentified.

The study is situated in the wider task-based exposure literature. The task-based approach to labour markets holds that technologies rarely automate whole occupations and instead alter jobs by displacing or augmenting particular tasks, which is why exposure is measured at the task level \parencite{Autor2003TheExploration,Acemoglu2018TheEmployment,Acemoglu2022TasksInequality}. Generative AI differs from earlier waves by targeting knowledge-intensive cognitive work that had resisted automation \parencite{Korinek2024TheAI,Autor2024ApplyingJobs}, and by diffusing, at least in part, bottom-up through worker discretion \parencite{Arntz2026barriers}. Exposure measurement has evolved with the technology being studied. \Textcite{Frey2017TheComputerisation}'s index of occupational-level automation-risk was highly influential despite measurement concerns \parencite{Evans2025MethodologicalWorkforce}. Their approach became the template for numerous subsequent automation-risk estimates \parencite{Filippi2023AutomationAgenda}. Machine-learning suitability scores assess whether tasks have the properties needed for machine learning \parencite{Brynjolfsson2017WhatImplications,Brynjolfsson2018WhatEconomy}; patent-derived measures link the direction of technological innovation to occupational tasks \parencite{Webb2020TheMarket}; and capability-based measures connect advances in particular AI applications to the abilities used in
occupations \parencite{Felten2021OccupationalUses,Felten2023OccupationalAI}. These measures
capture different technological margins. EMMR shifted the object of measurement to jobs' exposure to LLMs and showed that an LLM could serve as a task-exposure rater, reaching 66\% agreement with human annotations across 2,087 O*NET work activities. EMMR assessed validity through agreement between raters, face validity and correlations with existing indices, while also reporting that small prompt changes moved GPT-4's ratings.

Two criticisms of exposure-based designs have direct consequences for the current study. \textcite{lambert2026broken} show that LLM exposure does not isolate susceptibility to AI: it is bound up with the scope for remote work, which in their data accounts for the decline in early-career hiring instead. More broadly, rapid chatbot adoption may not yet have produced detectable effects on earnings or recorded hours \parencite{Humlum2025StillWaters}. We treat teleworkability and other digital tooling as confounds to be quantified, and report conditional associations. Furthermore, and upstream of the empirical application, \textcite{NBERw35110} show that even for a given prompt, LLM-generated exposure scores depend on the rating model and that the dependence carries through to substantive conclusions. We take that as the starting point and ask what follows for practice: which facets of the rating process actually move the measure, how to carry what remains into an estimate, and how to tell whether a given result can survive it.

The wider question of how far a finding depends on the model that produced its annotations is not confined to academic exposure indices. The UK Standard Skills Classification, released by Skills England in 2026, relies extensively on LLMs and text embeddings, but repeated runs of the same mapping prompts produced significantly different scores, leading to revised scoring procedures \parencite{SkillsEngland2026SSC}. LLMs are now routine research annotators, often matching expert coders \parencite{Gilardi2023ChatGPTTasks,Ziems2024CanScience,Gmyrek2024AUK,Manning2024AutomatedSubjects}. The LLM-as-judge literature has documented multiple caveats: sensitivity to prompt design, agreement that varies widely by task \parencite{Zheng2023JudgingArena,Bavaresco2024LLMsTasks}, and evaluators that favour their own outputs \parencite{Panickssery2024LLMOutputs}. Less attention has been paid to carrying that disagreement into downstream estimates and examining which conclusions survive it. Doing so is this paper's
main methodological contribution. The procedure applies wherever LLM-generated annotations enter a downstream statistical analysis.

The remainder is organised as follows. \Cref{s:data} describes the survey, vacancy and cross-national data. \Cref{s:gaisi_construction} sets out how the index is built and where measurement error can enter it. \Cref{s:properties} establishes how it behaves as an instrument, including what reproduces, what depends on design choices, and how disagreement is propagated. \Cref{s:application} applies it to exposure, pay and labour demand, and \cref{s:validity} assesses its predictive content and portability. \Cref{s:conclusion} concludes.

\section{Data}
\label{s:data}
The index is built from the British Skills and Employment Surveys (SES), a series of nationally representative repeated cross-sections of people in paid work in Britain, run about every five years since the mid-1980s \parencite{Butt2025SkillsReport,Felstead2019SES}. We use the 2017 and 2023--24 waves, which hold 3,304 and 5,784 workers with complete task data. The SES asks about job tasks, skills use, and workplace technologies with consistent wording across waves, which is what makes exposure measurable at the job rather than the occupation level \parencite{Henseke2025DegreesMarket}.

Since 1997, SES has asked workers to rate the importance of a consistent set of work activities or job tasks. Importance is reported on a 5-point scale from essential to not at all important. The instrument aims to collect information on the tasks performed in jobs across the economy to measure the demand for \textit{generic job skills} rather than to provide a complete account of activities workers carry out in their role \parencite{Green2012EmployeeAnalysis}. Its scope and scale are similar to O*NET's inventory of generalised work activities, which describes general job behaviours \parencite{Hansen2014ONETDevelopment}.

In SES~2017 and 2023--24, respondents rated up to 44 work activities, with some variation depending on managerial status and the importance of foundational literacy and numeracy tasks. \Cref{tab:taskcats} lists them, grouped into 12 categories for exposition only: every LLM classification operates on the individual task.

How a general job behaviour translates into concrete activity differs by occupation: selling or problem-solving involves different tools, workflows and settings in, for instance, software development than in a skilled trade. The ratings therefore condition on occupational context through vignettes built from 4-digit occupation unit group descriptions, at the 25 sub-major groups of the UK Standard Occupational Classification 2010 \parencite{ONS2010StandardGroups}. Workers report an average of 26 essential-equivalent tasks, with Collaboration ($M=0.80$), Problem Analysis ($M=0.73$) and Expertise ($M=0.71$) rated most important (\cref{tab:taskimportance}).

To follow pay across the period, we use the UK Quarterly Labour Force Survey (QLFS), pooling 2017~Q1 to 2026~Q1 \parencite{ONS2026QLFS}. The QLFS records no task content, so exposure enters it as an occupational average: each worker is assigned the mean exposure rank of their 3-digit SOC~2010 group, computed in SES~2017 and 2023--24. That gives 279{,}709 employees aged 20--65 across 89 occupation groups, and allows the pay gap between more and less exposed occupations to be tracked period by period over the whole window rather than at the two SES snapshots. Because the dose is an occupation average, these estimates carry no within-occupation task variation and are not on the same footing as the worker-level SES estimates.

To assess AI's association with labour demand, we merge GAISI scores into monthly counts of online job adverts summed to calendar quarters by 3-digit occupation and local authority district, published by the Office for National Statistics \parencite{ONS2025LabourUK}. Two series are used: the stock of live adverts, which ONS discontinued in July 2025, running from April 2017 to June 2025, and the flow of new postings, which runs to June 2026. Event-study specifications then compare hiring before and after ChatGPT's release in November 2022 within occupation-area cells.

To test whether GAISI travels beyond the UK and the SES task battery, we apply the same classification to the OECD Survey of Adult Skills (PIAAC) Cycle~2 \parencite{OECD2024Survey2023}. PIAAC collects worker-reported task data as SES does, but across countries and with different items and response scales, so agreement between the two indices is evidence that the approach does not depend on one instrument's wording.

\section{Construction of GAISI}
\label{s:gaisi_construction}
\subsection{Task-Level Scoring}
Building on EMMR, this section describes the process of scoring tasks within their occupational context and steps taken to aggregate those scores to the level of jobs. We classify job tasks' exposure to generative AI using commercial LLMs. The objective was to estimate the probability that generative AI could meaningfully augment each SES task within a specific occupational setting.

The classification distinguishes four levels of exposure:
\begin{itemize}
    \item \textbf{E0: No meaningful exposure.} LLMs do not reduce task time by $\geq$25\%
    \item \textbf{E1: Direct exposure.} LLMs alone (e.g., via a chatbot) reduce task time by $\geq$25\%.
    \item \textbf{E2: Latent exposure.} LLMs require integration with other tools or systems to achieve $\geq$25\% time savings.
    \item \textbf{E3: Exposure via image capabilities,} e.g., image recognition supports significant task efficiency.
\end{itemize}

We define the threshold of 25\% time savings for a typical competent worker, rather than 50\% as used in EMMR, as our criterion for meaningful exposure, consistent with assumptions of average labour cost savings \parencite{Acemoglu2025TheAI}. While this threshold is necessarily arbitrary, even smaller efficiency gains from LLM use may be meaningful in high-intensity work environments. In sensitivity checks, we assess how well the resulting index compares to versions using the original assessment rubric and cutpoints.

Unlike EMMR, which assigned tasks to mutually exclusive exposure categories, we prompted the LLMs to return a probability distribution across all four exposure levels (E0--E3). This decision reflects the high-level nature of SES work activities, which would typically comprise subtasks with varying susceptibility to AI. For example, a task like ``problem solving'' may include both elements amenable to generative AI augmentation or automation and others that are not. The probabilistic format is designed to accommodate cases where exposure-level distinctions are blurred. The prompt required that the probabilities sum to unity across exposure levels; ex-post verification confirmed compliance aside from minimal rounding errors across all 18,698 retrieved ratings (mean = 1.000, SD $\approx$ 0.000). Following EMMR, E2 and E3 were treated as latent exposure when constructing a summative index of generative AI exposure.

To capture the occupational context of tasks, we classified each of the 44 SES tasks separately for each of the 25 sub-major (2-digit) groups in the UK Standard Occupational Classification 2010, giving 1,100 task-occupation cells. Every cell is rated by nine raters drawn from three vendors and several model generations: Gemini~1.5~Pro, 2.5~Pro and 3.1~Pro; GPT-4o, 5.1, 5.4, 5.5 and 5.6-Luna; and GLM~5.2. The panel is a purposive selection, assembled so that the ratings vary by vendor and model generation, and, through the random seed, within a fixed configuration. It holds 18,698 ratings at the main prompt.\footnote{Two task-occupation ratings were lost due to API errors.} The panel is not intended to be representative of all LLMs, but rather to capture the range of plausible ratings that a researcher might obtain when using frontier LLMs as raters.

Each model was prompted using a structured rubric that guided the model to consider:
\begin{enumerate}
    \item The SES task's purpose and function within a UK-specific occupational vignette.
    \item The tools typically used to perform the task.
    \item The \textit{incremental contribution} of generative AI beyond existing technologies.
    \item A step-by-step justification of a probability distribution over exposure levels E0 to E3.
\end{enumerate}

As mentioned above, tasks were grouped into thematic categories (e.g., reading, collaboration) to reduce fragmentation and the number of API calls. Prompts also included classification examples and followed a chain-of-thought structure to encourage systematic reasoning. The rater temperature was set to 0.2 for Gemini~1.5~Pro and GPT-4o, which each ran five times over the full grid, giving 5,500 ratings apiece that are averaged before aggregation. A sixth run with GPT-4o switched sampling off (temperature = 0). The remaining seven raters were run once. Repeated classification of this kind establishes how stable the decoding process is; how far the ratings depend on the rater is a separate question, addressed by classifying the same cells across the panel. \Cref{ss:rating_precision} reports both. \Cref{app:A:prompt} reproduces the rater prompt.\footnote{The classification inputs, prompts, rater outputs and analysis code will be released as a replication package on publication.}

\subsection{Aggregation to the Job Level}
This study follows \textcite{Felten2021OccupationalUses} to combine task-level AI exposure into a job-level measure. First, we average each rater's ratings over its runs:

\begin{equation}
    P_{ok}^{e}(r) = \frac{1}{R_r} \sum_{j=1}^{R_r} x_{okj}^{e}(r)
\label{eq:eq1}
\end{equation}

In \cref{eq:eq1}, $x_{okj}^{e}(r)$ is the probability rater $r$ assigns to occupation-task $ok$ being of exposure level $e \in \{E0,E1,E2,E3\}$ on run $j$, and $R_r$ is the number of times that rater was run over the grid: five for Gemini~1.5~Pro and GPT-4o, one for each of the other seven. Averaging over runs removes decoding variance but not the rater. Every rater therefore has its own set of 1{,}100 average probabilities for each of the four exposure levels, and every step below is carried out once per rater.

The task-occupation probabilities are then merged into SES by occupation and task and weighted by how important the worker rates each task:

\begin{equation}
    E_i^{e}(r)=\frac{\sum_{k=1}^{44}P_{ok}^{e}(r) \times I_{ki}}
    {\sum_{k=1}^{44} I_{ki}}
\label{eq:eq2}
\end{equation}

The importance weight $I_{ki}$ runs from 1 where worker $i$ rates task $k$ as essential to 0 where the task plays no part in the job. Exposure at the job level therefore varies both with a worker's task profile, through $I_{ki}$, and with their occupation, through $P_{ok}^{e}(r)$. The denominator is the worker's total task load, the number of essential-equivalent tasks the job comprises, so that $E_i^{e}(r)$ is the share of that load falling in exposure level $e$. Without the adjustment, jobs made up of more essential tasks would score higher simply for having more of them.

\begin{figure}[p]
    \centering
    \caption{\FigOneCaption}
    \label{fig:fig1}
    \includegraphics[height=0.9\textheight]{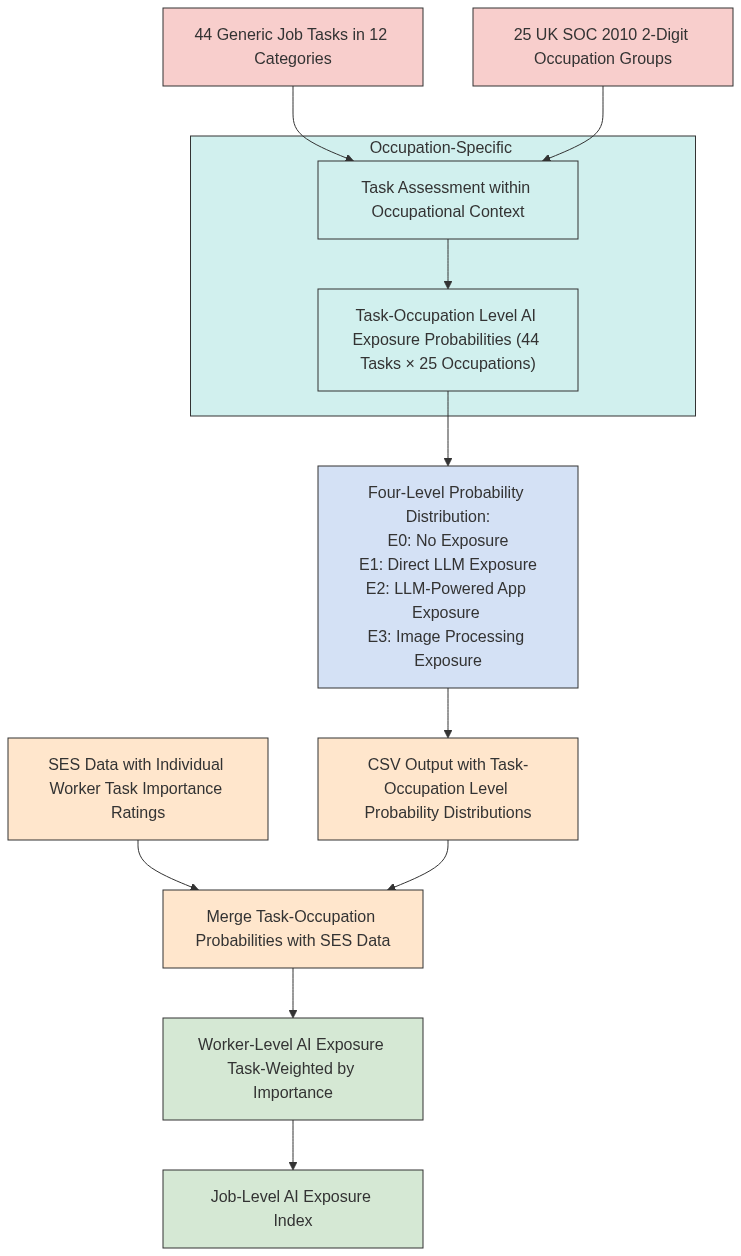}
\end{figure}

EMMR propose `discount' factors for combining latent exposure with direct exposure, weighting the latent components at 1 under full integration and at 0.5 under partial integration. We adopt their preferred 0.5 --- a mid-range scenario in which roughly half of the technically feasible integrations diffuse in the near term --- which gives the index

\begin{equation}
    G_i(r) = E_i^{E1}(r) + \omega \left[ E_i^{E2}(r) + E_i^{E3}(r) \right], \qquad \omega = 0.5
\label{eq:eq3}
\end{equation}

one GAISI score per worker per rater. \Cref{ss:robustness} compares $\omega = 0.5$ with the weight each rater's own ratings imply, and \cref{ss:rating_precision} sets out why the analyses that follow use a worker's rank in $G_i(r)$ rather than its level. \Cref{fig:fig1} visualises the classification pipeline.

\subsection{Sources of Measurement Error}
\label{ss:measurement_error}
Three sources of error can affect ratings produced this way, and they call for different responses.

Let $G_c^*$ denote the exposure the index is intended to capture for task-occupation cell $c$, and write the rating that rater $r$ returns for that cell as

\begin{equation}
    G_c(r) = G_c^* + b_c + \delta_c(r) + u_c(r)
\label{eq:eq4}
\end{equation}

The difference between $G_c(r)$ and $G_c^*$ need not be noise that averages out. $u_c(r)$ is idiosyncratic to the pairing of cell and rater and includes the decoding draw; $\delta_c(r)$ is a systematic deviation of rater $r$, shared by models of the same vendor or the same version; and $b_c$ is a possible deviation common to every rater. Averaging over runs shrinks $u$, and averaging over a panel shrinks $u$ and, as far as the panel spans vendors and versions, $\delta$. The design, however, does not touch $b_c$. If it contains bias rather than signal, averaging cannot reduce it, and the panel cannot distinguish it from $G_c^*$. 

$\delta$ and $u$ are the first error source. A different model or another decoding draw will return a different number. This part of the rating thus reflects choices made by the researcher, not something about the job. Because those choices are controlled and can be repeated, it can be estimated by rating the same cells many times under varied conditions, and then carried into downstream analyses, which is what this paper does.

A possible second source is the common deviation $b_c$. The training corpora of LLMs likely contain the academic and policy discourse that defines AI exposure in the first place, so the ratings may partly encode prevailing beliefs about what the technology can do rather than an independent assessment of it. If the models share much of that corpus, and if the resulting common component is mismeasurement rather than a valid assessment, repeating the exercise across them provides little purchase. Because $b_c$ and $G_c^*$ enter identically for every rater, the panel cannot establish whether $b_c$ is non-zero or estimate its size; if present, it falls inside what a variance decomposition attributes to the cell (\cref{ss:rating_precision}).

This possibility creates a specific difficulty for validation. Let $A^*$ be a worker's actual use of AI at work and $A = A^* + e_2$ a self-reported measure. Bias from circularity requires both a non-zero common deviation $b$ and a reporting error $e_2$ that is positively associated with it. If public discourse leads raters to overstate AI capability for salient activities and workers to be more likely to label software supporting those activities as AI, this could yield $cov(b,e_2)>0$, so that the observed association between exposure and reported use is confounded by the narrative common to both. Discussion of AI-assisted problem solving, for example, could raise both a rater's assessment and a worker's propensity to recognise problem-solving software as AI, while less salient assistance such as grammar correction goes unrecognised. Neither the existence of $b$ nor its covariance with $e_2$ is identified. It is $b$, rather than $\delta$ or $u$, that matters for this conjecture, since a worker's reporting error has no reason to track which model was asked. The indirect evidence below asks whether the index carries information beyond this possible mechanism; it cannot test whether $b=0$.

If the association between exposure and reported use were largely narrative, the index should add little to existing technology exposure measures, which are built from the same literature. It should be weaker for a survey question that does not mention AI than for one that does. Moreover, pay and vacancies do not share the same susceptibility to AI discourse. They can be informative about circularity to the extent that exposure-based research designs separate exposure from competing occupational trends. None of these tests is decisive alone; together they shed light on whether the index is relevant despite potentially systematic measurement error. \Cref{ss:circularity} returns to this point.

A third source of error is specific to a panel that spans model generations. The nine raters were released between 2024 and 2026, so the later vintages were trained on a corpus that already recorded how the technology was being used. A rating may then register not what a language model could do for a task but what language models were by then observed doing, one of the mechanisms \textcite{NBERw35110} propose. This operates through $A^*$ rather than through reporting error, making $\delta_c(r)$ depend on use itself, and because $\delta$ varies with vintage, a panel spanning model generations can test it. If raters calibrate on diffusion, later vintages should rate the tasks that had come into use as more exposed, so measured exposure should rise with model generation, and rise most where use concentrated. \Cref{ss:vintage} tests this directly.

Two design responses follow from these challenges. Ratings come from a panel of raters rather than a single model, so that $\delta$ and $u$ are estimated and propagated (\cref{s:properties}). Further, the index is validated against self-reports as well as against outcomes that do not pass through worker perceptions of AI, namely job postings and wages, which record employer decisions and market prices (\cref{s:application,s:validity}). 

\section{Properties of the Index}
\label{s:properties}

Exposure to generative AI is not directly observable, so GAISI is judged as an instrument: how reproducibly it ranks jobs when the rating process is varied, how far it depends on choices made in its construction, and how the remaining rater disagreement is carried into the analysis. This section reviews those properties before the index is applied in \cref{s:application}; \cref{s:validity} then asks whether it measures what it is intended to.

\subsection{Reproducibility of the Ratings}
\label{ss:rating_precision}

How reproducible an LLM rating is depends on what parameter one varies. Exposure scores are ordinarily produced by a single model in a single pass. \Textcite{NBERw35110} show that substituting one frontier model for another moves the scores and the conclusions drawn from them. Their finding raises a broader question: among the several things that can differ between one rating and another, including the random seed, the wording of the prompt, the model, or the vendor, which of them actually move the measure? \Cref{tab:raterreliability} answers it by estimating reliability separately for each rater in the panel, resampling one feature of the rating process at a time and holding the rest fixed. Reliability here is the consistency intraclass correlation: how alike two ratings order the 1,100 task-occupation cells, setting aside any constant difference in level between them. 

The ordering in \cref{tab:raterreliability} is the result. Running the same model on the same prompt, so that only the random seed differs, leaves the ranking almost intact: a single rating recovers it at 0.962 for Gemini~1.5~Pro and 0.891 for GPT-4o, the two models rated five times each. A grid rated at temperature 0 with GPT-4o, where nothing is sampled, confirms as much. Its ordering of the cells agrees with each of the five sampled runs at $\rho = 0.91$ on average, while those runs agree with one another at $\rho = 0.90$. Switching the sampling off therefore neither shifts the ratings nor materially alters how consistently they rank the cells.

Other changes are more consequential. Altering the prompt slightly (simplifying the wording, cutting back the worked examples, or removing the reference to existing workplace technology) lowers consistency to 0.845, and moving between model versions within one vendor's lineage lowers it to 0.841 for OpenAI and 0.745 for Google. A rating from one vendor recovers another vendor's ranking at 0.756. The pattern is thus not simply that vendors differ: version drift within Google's own vintages (0.745) is as large as the gap between vendors (0.756). A panel assembled to span vendors but not model generations would understate how far the ranking moves.

\begin{table}[htbp]
\centering
\caption{Reliability of Task--Occupation Exposure Ratings by Source of Variation} \label{tab:raterreliability}
\begin{threeparttable}
\begin{tabularx}{\textwidth}{l c *{2}{>{\centering\arraybackslash}X}}
\toprule
\textbf{Source of variation} & \textbf{Raters} & \textbf{Single rater} & \textbf{Ensemble} \\
\midrule
Random seed (Gemini~1.5~Pro)     & 5 & 0.962 & 0.992 \\
Random seed (GPT-4o)             & 5 & 0.891 & 0.976 \\
Prompt wording (Gemini~2.5~Pro)  & 4 & 0.845 & 0.956 \\
Model version, within OpenAI     & 5 & 0.841 & 0.963 \\
Model version, within Google     & 3 & 0.745 & 0.897 \\
Vendor                           & 3 & 0.756 & 0.903 \\
\bottomrule
\end{tabularx}
\begin{tablenotes}[flushleft]
\footnotesize
\item \textit{Note:} Each row reports how reliably the 1,100 task--occupation exposure ratings are reproduced when one feature of the rating process is varied, and the others are held fixed. Entries are consistency intraclass correlations, which measure the reliability of the \textit{ranking} of cells; higher values mean the raters order the cells more alike. ``Single rater'' is the reliability of one rating drawn from that source; ``Ensemble'' is the reliability of the average of the number of raters shown. Each row is estimated on the subset of the rater panel that identifies it, restricted to cells rated by every rater in that subset. Every row here varies how the same quantity is elicited. The four prompt variants (the main wording, simplified phrasing, fewer worked examples, and no reference to existing workplace technology) request a probability distribution over the same four exposure levels at the same 25\% threshold. The prompt row is estimated on a single model, so it is not a general prompt effect. The vendor row compares one model from each of three vendors.
\end{tablenotes}
\end{threeparttable}
\end{table}

Taking the nine raters together, three-quarters of the variance in the ratings separates task-occupation cells and one quarter separates raters (Appendix \cref{tab:appB:raterpanel}). Of the rater quarter, 11\% is attributable to vendor style and 17\% to model version within a vendor --- together the systematic rater component $\delta$ of \cref{eq:eq4} --- leaving 72\% idiosyncratic to the particular pairing of cell and rater, which is $u$. The nine models therefore behave more like nine imperfectly correlated annotators than like three vendor blocs, and averaging over them pays: the share of variance that distinguishes cells is 0.75 for a single rating compared to 0.96 for the average of all nine. However, the decomposition cannot separate $G^*$ from $b$. If present, a shared bias would enter every rating identically, so it would sit inside the three-quarters that distinguishes cells. The between-cell share therefore cannot be interpreted as pure signal, and the ensemble figure of 0.96 (Appendix \cref{tab:appB:raterpanel}, Panel~B) is reliability with respect to $\delta$ and $u$ alone. 

There is a clear asymmetry in interrater agreement between ranks and levels. Pairwise rank correlations between the nine raters run from 0.74 to 0.92, with a median of 0.81. By contrast, comparisons of exposure levels do not hold up: the share of British jobs placed above an exposure of 0.5 ranges from under 0.1\% to 38\% across raters, and mean exposure from 0.29 to 0.46, even though every rater assigns all jobs some exposure. Which jobs are more AI exposed than others is consistent and reproducible, whereas how many jobs count as ``highly exposed'' is a matter of the rater's calibration. Exposure therefore enters every analysis below as a percentile rank. Every substantive estimate is reported once per rater and pooled across raters. 

Raters' scores are also internally consistent with the generated justifications of their ratings: affordance cues predict higher scores, while human constraints and explicit limitations predict lower ones, including within task--occupation cells (\cref{tab:appA:grounding}). Coding the justifications with a second model produces similar results.

\subsection{Sensitivity to Design Choices}
\label{ss:robustness}

Four design elements have consequences for the subsequent analysis: the share of working time the SES battery observes, the rubric put to the rater, the 25\% threshold that rubric encodes, and the weight on latent exposure (\cref{s:gaisi_construction}). We assess each against reported AI use at the worker level in SES~2023--24 across all nine raters. Discrimination is measured by the area under the receiver operating characteristic curve, where 0.5 is chance, and 0.70 serves as a pragmatic benchmark for useful predictive power.

\textbf{What the task battery measures.} The 44 SES activities account for about a third of working hours, ranging from 17\% to 46\% across occupational groups (\cref{tab:appB:coverage_soc1}). The index normalises over the activities observed in the battery, effectively treating them as representative of the omitted remainder. We compare this construction with two extreme alternatives, assigning either no exposure or full exposure to all working time outside the battery.

The index as constructed predicts reported AI use with an AUC of 0.72. Discrimination falls to 0.66 when all omitted time is treated as AI-resilient and to 0.52, close to chance, when it is treated as fully exposed; both differences are significant under every rater (\cref{tab:appB:coverage_bounds}). The pattern is consistent with omitted working time comprising a mixture of more and less exposed activities, as the observed task battery does, rather than lying uniformly at either end of the exposure scale.

\textbf{Rubric.} \Cref{tab:raterreliability} varies how the same quantity is elicited. What happens when the quantity itself changes? EMMR's original rubric asks for a single category rather than a distribution: it agrees with the same annotator's own ratings at 0.57, below any pair in the nine-rater panel, and predicts AI use at 0.685 against the benchmark's 0.725 ($p = 0.004$). It is also the only variant whose level sits outside the panel's range, at a weighted exposure mean of 0.54 against 0.29 to 0.46 across the nine raters (\cref{tab:appB:prompt_permutation}). What is asked for thus moves the predictive power in this set of LLM-generated ratings.

\textbf{Threshold.} Raising the bar from a quarter to a half of task time, holding the rest of the prompt fixed, changes the level statistic and leaves the ordering intact. The share of jobs scoring above 0.5 falls from 29\% to 4\%, while rank agreement with the baseline thresholds of 25\% is 0.90 across task-occupation cells and 0.91 across workers. The predicted change in reported AI use is 0.190 per interquartile range at the 25\% threshold against 0.204 at 50\% (standard errors 0.022 and 0.025). The threshold governs how many jobs are called exposed, not which ones.

\textbf{Weight on latent exposure.} The weight the ratings imply for latent exposure is below the conventional 0.5 for most raters, but no result turns on the difference. Entering the direct and latent components separately in a model of worker-reported AI use, both as percentile ranks, and taking the ratio of their coefficients gives 0.335 (SE 0.127) for the benchmark rater and 0.330 (SE 0.134) for Gemini~2.5~Pro; six of the nine intervals contain 0.5. Reported use responds more strongly to the direct component than to the latent one for eight of the nine raters. The exception is GLM~5.2, which reverses the ordering and returns 2.04; excluding it, the ratio runs from $-0.08$ to 0.73 with a mean of 0.33. Prediction is indifferent to the choice: an index built on each rater's implied weight and one built on 0.5 differ by less than 0.01 AUC for every rater (\cref{tab:appB:e2_weight}). The 0.5 weight is a convention, retained for comparability with the EMMR-based literature.

\subsection{Carrying Rater Disagreement into the Analysis}
\label{ss:propagation}

Because no single rater is authoritative, every substantive analysis is estimated nine times, once under each rater's version of the index. The rater-specific estimates and their minimum--maximum range are the primary
evidence on sensitivity to model choice. Reporting this set constitutes a multiverse analysis in the sense of
\textcite{Steegen2016IncreasingAnalysis} and \textcite{Simonsohn2020SpecificationDescriptions}, with the rating model as the analytic choice being varied.

For a compact summary, we also average the nine coefficients and use Rubin's variance-combination formula \parencite{Rubin1987MultipleSurveys} as a pragmatic accounting device. The resulting variance combines the average sampling variance within raters with the observed dispersion of the coefficients between raters. Its purpose is to prevent rater disagreement from disappearing when the results are reduced to a single estimate.

This is a heuristic use of the formula. The nine indices are not missing-data imputations, posterior draws, or a random sample from a defined population of LLMs. The resulting interval therefore does not have a formal coverage interpretation over all possible raters. It summarises sampling uncertainty and disagreement among the nine raters considered here. It captures variation associated with the rater-specific components $\delta$ and $u$ only insofar as that variation appears in the panel; it cannot reveal whether a shared component $b$ exists, or its size if it does, and it cannot capture sensitivity to models and design choices outside the panel. We therefore report the rater-specific range alongside every pooled summary.

Within this Rubin-style accounting, the relative importance of rater disagreement will depend on the precision of the downstream design. When an estimate is precise within each rater, the observed dispersion between
rater-specific coefficients forms a larger share of the combined variance; when sampling uncertainty is substantial, it forms a smaller share. This share is therefore a descriptive diagnostic for the nine-rater multiverse, not an estimate of the proportion of all measurement error caused by rater choice. The two applications below differ sharply on this diagnostic, and we report it alongside the rater-specific estimates and their range.

\section{Assessing the Reach of Generative AI in Britain's Workforce}
\label{s:application}
We apply GAISI to worker-level data from the SES and the Labour Force Survey, and to online job vacancies measured at occupation~$\times$~area level. The index is built nine times over, once from each rater in the panel, and every estimate below is reported once per rater and pooled across them (\cref{ss:propagation}). The following maps exposure across tasks and occupations, describes who holds the most exposed jobs and how far those workers report using AI, and relates exposure to pay and to the demand for labour.

\subsection{Mapping AI Exposure in Work Tasks and Jobs}
\label{ss:mapping}
Exposure rises with occupational skill requirements, though only coarsely. \Cref{tab:gaisi_occ} reports, for each occupational skill group, the average position its workers occupy in the exposure ordering. Higher managerial, professional, and technical workers sit at the 69th percentile of that ordering and lower managerial, associate professional, and clerical workers at the 65th, against the 29th percentile for skilled trades, personal services, and sales, and the 14th for operatives, elementary, and customer service occupations. The two upper groups sit 37 percentile points above the two lower ones, a gap no rater puts below 31 points. The four-point difference between the two upper groups is less clear-cut: two of the nine raters reverse it by less than a percentile point. In all, the exposure gradient runs opposite to earlier waves of automating technology, which concentrated on routine manual and clerical work \parencite{autor2013growth}.

The gradient is inherited from the task ratings. Writing and reading (long) documents, keeping up to date and applying new knowledge are the four activities the panel places highest of the 44 in the SES battery: none falls below the 19th under any rater, and three never leave the top eight. Physical strength and physical stamina sit at the other end, ranked 43rd and 44th by every rater. Jobs built around long-form reading and writing are concentrated among professional and managerial occupations, and the occupation skill gradient follows from that concentration.

\Cref{tab:gaisi_occ} also lists the occupations at each end. Research and development managers, information technology and telecommunications professionals, and media professionals rank highest of the 78 three-digit occupations with more than ten sample members, and elementary storage, process plant and construction occupations rank lowest. The bottom three are firmly placed: no rater lifts any of them out of the lowest eight. The top three are stable in kind but not in position, with media professionals ranging from 1st to 15th across the panel. Four of the six named occupations are measured on fewer than 40 sample members, which is partially why their positions move.

\begin{table}[htbp]
\centering
\begin{threeparttable}
\caption{\TabFiveCaption}
\label{tab:gaisi_occ}
\begin{tabularx}{\textwidth}{lc *{2}{>{\centering\arraybackslash}X}}
\toprule
 & \textbf{N} & \textbf{Exposure percentile} & \textbf{Range across raters} \\
\midrule
\multicolumn{4}{l}{\textit{Panel A. Occupation skill level}} \\
High (Higher Managers, Professionals, Technicians)        & 2,158 & 69.0 & 66.5--71.7 \\
Mid-High (Lower Managers, Assoc Professionals, Clerical)  & 1,644 & 65.3 & 62.2--68.2 \\
Mid-Low (Skilled Trades, Personal Services, Sales)        & 915   & 28.7 & 27.0--30.6 \\
Low (Operatives, Elementary, Customer Service)            & 1,067 & 13.9 & 12.0--15.3 \\
\midrule
\multicolumn{4}{l}{\textit{Panel B. Three highest-ranked three-digit occupations}} \\
 & \textbf{N} & \textbf{Position in ordering} & \textbf{Range across raters} \\
Research and Development Managers                            & 18  & 1 & 1--11 \\
Information Technology and Telecommunications Professionals  & 260 & 2 & 1--12 \\
Media Professionals                                          & 34  & 3 & 1--15 \\
\midrule
\multicolumn{4}{l}{\textit{Panel C. Three lowest-ranked three-digit occupations}} \\
Elementary Storage Occupations        & 71 & 76 & 71--77 \\
Elementary Process Plant Occupations  & 22 & 77 & 73--78 \\
Elementary Construction Occupations   & 14 & 78 & 72--78 \\
\bottomrule
\end{tabularx}
\begin{tablenotes}[flushleft]
\footnotesize
\item \textit{Note:} Each of the 5,784 workers in the SES~2023--24 sample is placed at their percentile in the exposure distribution, on a 0--100 scale, and the percentile is recomputed separately for each of the nine raters in the panel. Panel~A gives the weighted mean percentile of workers in each occupational skill group. Panels~B and~C give the position of individual three-digit occupations in an ordering of 78 such occupations, 1 being the most exposed, where occupations are ordered by the mean percentile of their workers averaged over the nine raters. ``Range across raters'' is the minimum and maximum over the nine raters of the quantity in the preceding column. Seven of the nine raters place the High skill group above Mid-High; all nine place Mid-High above Mid-Low and Mid-Low above Low. Skill levels are grouped by two-digit SOC~2010 codes: High (11, 21, 22, 23, 24, 31), Mid-High (12, 32, 33, 34, 35, 41, 42), Mid-Low (52, 54, 61, 62, 72) and Low (51, 53, 71, 81, 82, 91, 92). Three-digit occupations with ten or fewer sample members are excluded; the smallest cell shown holds 14 workers. Thin cells are partially why an occupation's position moves by several places between raters. 
\item \textit{Source:} SES~2023--24 (UK). Survey weights applied.
\end{tablenotes}
\end{threeparttable}
\end{table}

\subsection{AI Exposure and Use across the Workforce}
\label{ss:workforce}
Where a worker sits in the exposure ordering is mostly about the job, not the individual. Women sit 0.04 percentile points above men, indistinguishable from zero ($p = 0.99$) and never more than two points either way across the nine raters. The gap between white and ethnic minority workers is 3.7 points ($p = 0.056$). Age differences are more pronounced: those in their thirties, forties and fifties sit between 6.9 and 7.8 percentile points above workers aged 20--29, and those aged 60--65 sit 5.0 points above them. The education gradient is strongest. Relative to workers whose highest qualification is at secondary level, graduates sit 25.3 percentile points higher in the exposure ordering and those below secondary level 11.9 points lower --- a spread of 37 points, comparable to the gap between the top and bottom occupational skill groups in \cref{tab:gaisi_occ}.

Occupation absorbs nearly all of this. Conditioning on three-digit occupation removes at least 86 per cent of each gap that is significant to begin with, and no less than 65 per cent under the least favourable rater. Once occupation is absorbed, only the graduate `premium' remains distinguishable from zero, at 2.7 percentile points (SE 0.6); the four age terms and the ethnicity gap do not (results not shown separately).
 
\begin{figure}[htbp]
    \centering
    \caption{\FigFourCaption}
    \label{fig:fig4}
    \includegraphics[width=0.9\textwidth]{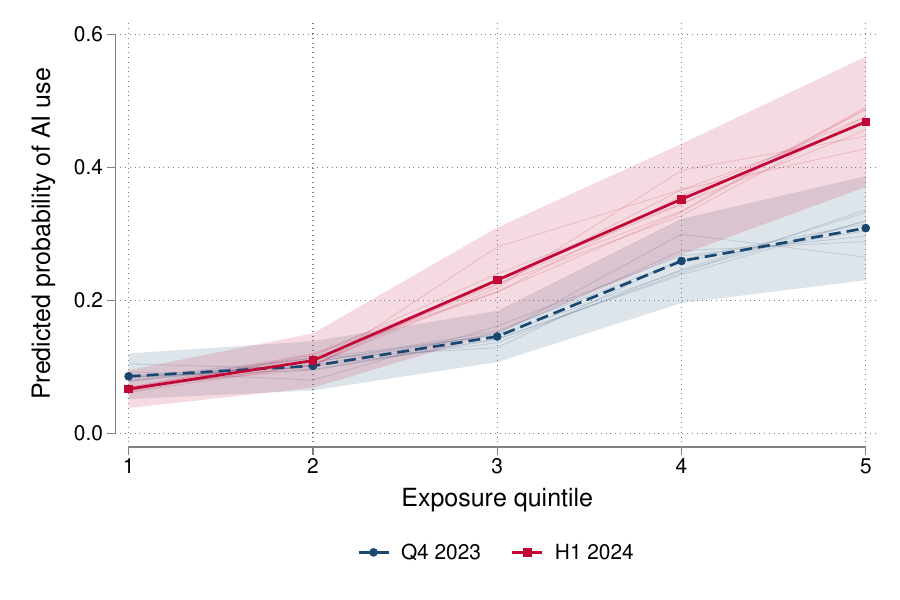}
    \begin{minipage}{0.9\textwidth}
    \footnotesize
    \textit{Note:} Predicted probability of reporting use of AI-powered software at work, by quintile of the worker's exposure percentile, shown separately for the two SES~2023--24 fieldwork periods (Q4~2023 and H1~2024). Estimates come from a probit model controlling for age group, sex, education, ethnicity and region, weighted for survey design and non-response, with standard errors clustered by four-digit occupation. The model is estimated once for each of the nine raters in the panel, with quintiles cut on that rater's own exposure ranking; the solid lines give the estimates combined by the Rubin-style pooling and the shaded areas their 95\% intervals, which carry both sampling variation and disagreement between raters. The nine rater-specific series are drawn faintly behind them. \textit{Source:} SES~2023--24 (UK), ages 20--65. $N = 5{,}708$. Authors' calculations.
    \end{minipage}
\end{figure}

Reported use of AI at work rises with exposure, and the increase in use between the start and end of the fieldwork periods was confined to the more exposed jobs. Across workers aged 20--65, the share reporting use of AI-powered software rose from 18 per cent in the fourth quarter of 2023 to 25 per cent in the first half of 2024. \Cref{fig:fig4} splits that by exposure quintile, holding age, sex, education, ethnicity and region fixed. Predicted use in the top quintile rose from 0.31 to 0.47, while in the bottom quintile it did not rise at all (0.09 to 0.07). Growth was restricted to the upper three quintiles: within-quintile increases of 0.09, 0.09 and 0.16 in the third, fourth and fifth are each distinguishable from zero, while the lowest two are not. The 2024 profile rises monotonically with exposure under all nine raters and the 2023 profile under six of them, and the band around the pooled line in \cref{fig:fig4} shows how far the picture moves with the choice of rater. Worker-reported adoption is following the exposure ordering rather than spreading evenly across jobs.

\subsection{Exposure and Labour Market Gradients, 2017--2026}
\label{ss:macro_signals}

\subsubsection*{Change in AI Exposure since 2017}
\begin{figure}[htbp]
    \centering
    \caption{\FigFiveCaption}
    \label{fig:fig5}
    \includegraphics[width=0.9\textwidth]{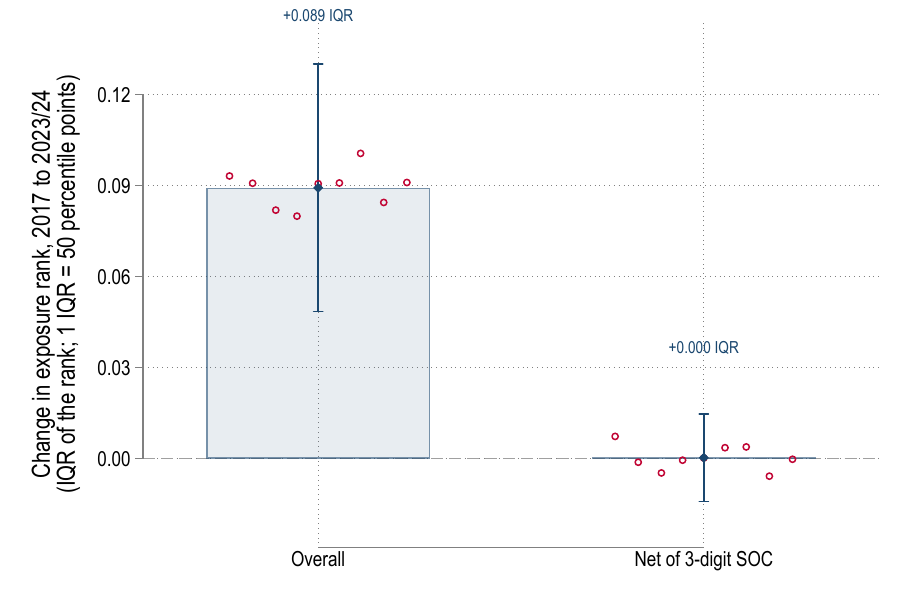}
    \begin{minipage}{0.9\textwidth}
    \footnotesize
    \textit{Note:} Bars give the change in mean worker exposure rank pooled across the nine raters using a Rubin-style approach; whiskers are 95\% confidence intervals and open circles are the rater-specific estimates. The second estimate absorbs 3-digit SOC~2010 fixed effects, leaving the within-occupation margin. Exposure ranks are cut once over the pooled SES~2017 and 2023--24 samples. \textit{Source:} SES~2017 and 2023--24 (GB), $N=8{,}167$ workers; authors' calculations.
    \end{minipage}
\end{figure}

Aggregate exposure rose between 2017 and 2023/24, and essentially all of the rise is compositional. \Cref{fig:fig5} plots both margins. Mean exposure rank rose by 0.089 IQR over the two waves (Rubin-pooled, SE 0.021, $p < 0.001$), or 4.5 percentile points, and every rater puts the rise between 0.080 and 0.101 IQR. Net of 3-digit occupation, the same change is $\approx$0.000 IQR (SE 0.007, $p = 0.96$), and not significantly different from zero under any rater. The within-occupation margin accounts for 0.4\% of the total movement and the between-occupation margin for 99.6\%. The average worker in 2023/24 holds a more exposed job than in 2017; they do not report a more exposed mix of tasks within a given job. The occupations that grew are the ones that ranked high on exposure in 2017: the employment-weighted rank correlation between an occupation's 2017 exposure rank and its subsequent change in employment share is 0.55 (Fisher-$z$ pooled across 86 occupations), between 0.50 and 0.61 across raters and significant under all nine.

\subsubsection*{Pay and Exposure, 2017--2026}
The pay gap between more and less exposed occupations has been narrowing since 2017, and it has continued to narrow after ChatGPT without breaking trend around its launch. We estimate this in the UK Quarterly Labour Force Survey: 279{,}709 employees aged 20--65 across 2017 to 2026~Q1 with reported hourly pay, in 89 three-digit SOC 2010 groups. We regress log hourly pay on interactions between an occupation's mean exposure percentile rank, taken from SES, and four period indicators --- pre-pandemic (2017--19), pandemic (2020--21), post-pandemic (2022) and post-ChatGPT (2023 to 2026~Q1) --- with age, sex, education and country of birth as controls, three-digit occupation absorbed, and standard errors clustered on occupation. Absorbing occupation removes the level of the exposure premium along with every other time-invariant occupational characteristic; what is left is whether the gap moved. Coefficients are per interquartile range of the exposure ranking, as elsewhere in this section. One IQR is close to the spread in the LFS: the employment-weighted interquartile distance between three-digit occupations is 52 percentile points, or 1.05 IQR, so a per-IQR coefficient roughly corresponds to the gap between an occupation at the 29th percentile of exposure and one at the 81st. 

The year-by-year path shows a gradient that was moving well before generative AI. Against 2022, the interaction runs $+0.031$, $+0.033$, $+0.023$ and $+0.022$ for 2017 through 2020, reaches zero in 2021 ($-0.001$), holds there in 2023 ($+0.003$, $p = 0.74$), and then falls again: $-0.017$ in 2024 ($p = 0.037$), $-0.037$ in 2025 and $-0.046$ in 2026~Q1 (both $p < 0.001$).\footnote{Estimated on the benchmark rating; the period contrasts in \cref{tab:appC:ai_price_diffs} carry the full panel.} Parallel trends thus fail: the gap was wider in 2017--19 than in 2022 by 0.027 log points pooled ($p = 0.0005$, positive and significant under all nine raters), while 2020--21 is indistinguishable from 2022 ($0.009$, $p = 0.21$, no rater significant). A joint test that both pre-treatment interactions are zero gives $F(2,88) = 11.51$, $p < 0.0001$, and equality of the two pre-periods is rejected at $p = 0.0002$. A decline that resumes in 2024 rather than in the launch quarter fits the diffusion of the technology better than movement around ChatGPT's release date, but it is not an identified effect.\footnote{Within the SES, the same comparison at worker-level points the same way but remains underpowered (\cref{tab:appC:ai_price_diffs}, Panel B).}

Against the 2022 reference, pay in 2023 to 2026~Q1 is 0.020 log points lower per IQR of exposure, pooling the nine raters by the Rubin-style approach (pooled SE 0.007, $p = 0.005$). Every rater gives a negative estimate significant at the 5\% level, between $-0.018$ and $-0.023$, or 1.8\% to 2.2\% of hourly pay (\cref{tab:appC:ai_price_diffs}, Panel A). The between-rater share of the Rubin-combined variance accounts for 0.047 of the uncertainty around the pooled estimate, and sampling variation for the other nineteen-twentieths. Which model in the panel supplied the exposure score matters relatively little here.


\subsubsection*{Labour Demand for AI-Exposed Jobs}
Online job adverts in more exposed occupations have fallen relative to less exposed occupations since the release of ChatGPT, in series published by the Office for National Statistics. In the stock of live adverts, a difference of one interquartile range in an occupation's exposure rank is associated with 0.175 log points fewer adverts over the ten quarters from 2023 Q1 to 2025 Q2 (Rubin-pooled, SE 0.017, $p < 0.001$), or 10.0\% fewer per thirty percentile points of exposure. In the flow of new postings, the same contrast is 0.147 log points (SE 0.031, $p = 0.001$), 8.4\% per thirty percentile points, and it remains at 0.128 log points ($p = 0.007$) when the window is extended through 2026 Q2. Each of the nine raters gives a negative estimate significant at the 0.1\% level in both series, between $-0.158$ and $-0.193$ in the stock and between $-0.117$ and $-0.185$ in the flow (\cref{tab:appC:ai_demand_diffs}).

\begin{figure}[htbp]
    \centering
    \caption{\FigSixCaption}
    \label{fig:fig6}
    \includegraphics[width=\textwidth]{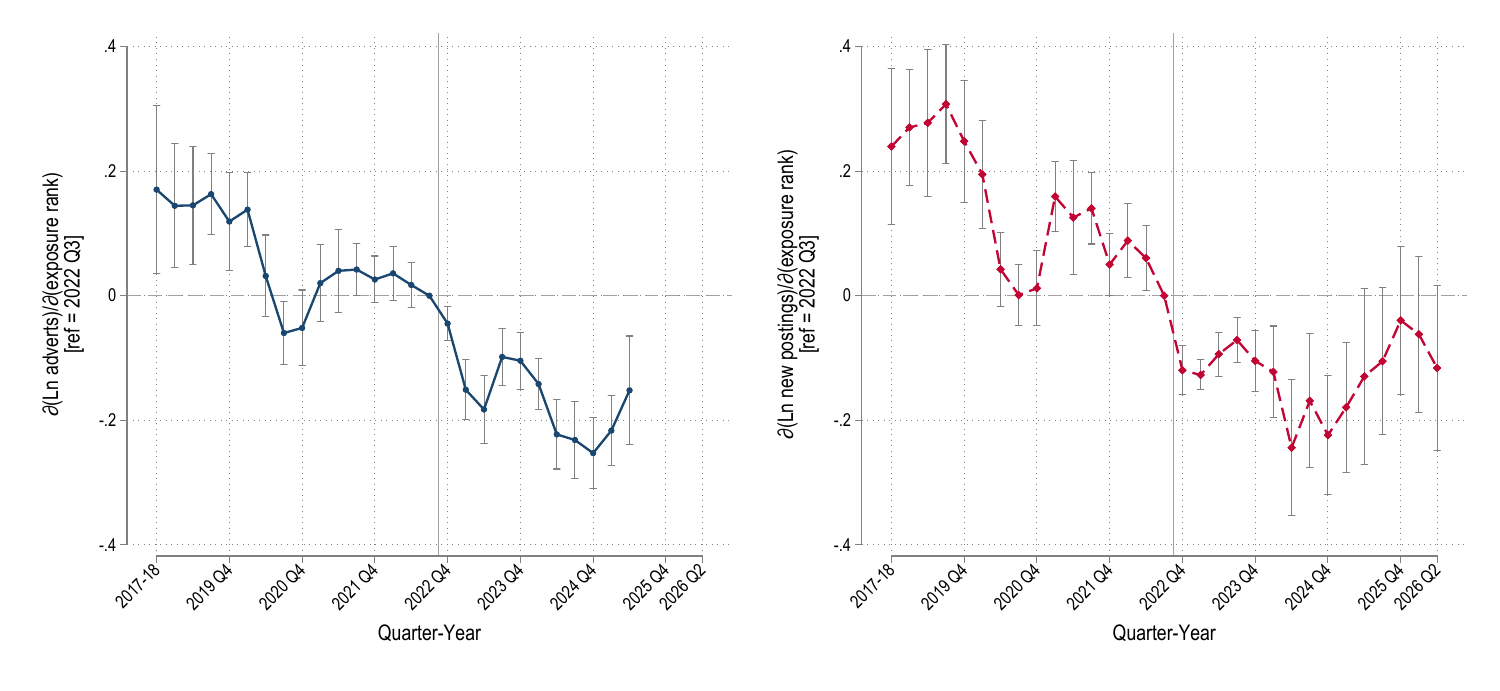}

    \par\smallskip
    \begin{minipage}{\textwidth}
    \footnotesize
    \textit{Notes:} Each point is the Rubin-pooled coefficient on the interaction between occupational generative-AI exposure and calendar quarter; whiskers show 95\% confidence intervals. Exposure is the occupation's mean percentile rank, scaled so that one unit equals one interquartile range. The reference quarter is 2022 Q3, and pre-2019 quarters are binned. The left panel is the stock of live adverts, which ends in 2025 Q2 because the ONS series was discontinued; the right panel is the flow of new postings, which runs through 2026 Q2. Models absorb occupation--local-area, region--quarter, and one-digit-occupation--quarter fixed effects; standard errors are clustered by occupation--local area. \textit{Source:} ONS Labour demand volumes by SOC; authors' calculations.
    \end{minipage}
\end{figure}

We merge each occupation's mean exposure percentile rank into ONS counts of online job adverts at the three-digit SOC 2020 by local authority district by quarter level, and regress the log of adverts plus one on that rank interacted with quarter dummies, taking 2022 Q3 --- the quarter before ChatGPT's release --- as the reference. Fixed effects for three-digit occupation $\times$ local area, one-digit occupation $\times$ quarter, and region $\times$ quarter absorb movements common to a broad occupational group or to a region, so the comparison is between more and less exposed occupations facing the same conditions; standard errors are clustered on occupation $\times$ local area. The two series measure different outcomes: a flow responds faster and with less persistence than a stock, and the two paths in \cref{fig:fig6} differ accordingly. The stock panel ends in 2025 Q2 (1,124,376 cells) and the flow panel in 2026 Q2 (1,261,219 cells).\footnote{ONS discontinued the stock of live adverts after its July 2025 release and has counted new postings only since. June 2026 is set aside because ONS suppressed 30\% of its cells that month.}

Similar to pay, the relationship between exposure and hiring was already moving before ChatGPT. Averaged over 2017 Q2 to 2022 Q2, the pre-period coefficient is $+0.096$ in the stock series (SE 0.033, $p = 0.016$) and $+0.174$ in the flow (SE 0.036, $p = 0.001$), positive under all nine raters in both. The event-time path in \cref{fig:fig6} locates that movement more precisely. In the flow series, the differential runs between $+0.24$ and $+0.31$ across 2017--19, and by 2020 Q3 it is $+0.001$ with a 95\% interval of $[-0.048, 0.050]$: tight enough to rule out a pre-period gap much beyond a third of the post-GPT release one. More exposed occupations were posting relatively more in the late 2010s, that gap closed over the pandemic, and the post-2022 coefficients are a break in a series that had been trending rather than a departure from a flat baseline. The design does not support parallel pre-trends, and the post-ChatGPT estimates are correspondingly a description of when the series turned.

Teleworkability is where the result breaks. Exposure and teleworkability are nearly collinear across occupations (their correlation at three-digit SOC 2020 is 0.82), so the question is whether the post-ChatGPT movement in postings can be attributed to exposure at all. It cannot. Collapsing the quarterly path into the same four broad periods used for pay, the post-2022 differential in new postings is $-0.137$ per IQR (SE 0.033, $p = 0.003$), negative and significant under all nine raters. Adding teleworkability --- measured from the 2021--22 Annual Population Survey, and given its own period path --- moves the pooled estimate to $+0.094$ (SE 0.069, $p = 0.21$), and the between-rater share of the Rubin-combined variance rises to 0.98 (\cref{tab:appC:vacancy_telework}). \footnote{The teleworkability measure is the share of an occupation's full-time workers mainly working from home, built natively on SOC 2020 from the Annual Population Survey of July 2021 to June 2022.}

The rater-multiverse provides further insights. Seven of the nine raters return individually significant positive post-2022 coefficients under the teleworkability control, running as high as $+0.167$, while Gemini~2.5~Pro returns $-0.015$ and GPT-4o $-0.008$. A single-rater study could therefore have reported either a precise positive association or no association, depending on which model supplied the ratings. In the pay analysis, the rater-specific estimates cluster tightly and between-rater dispersion forms only a small part of the Rubin-combined variance. The sensitivity of a downstream estimate to rater disagreement therefore depends on the independent variation left by the research design.

Job adverts are administrative records of employer decisions, and not subject to self-reporting bias (\cref{ss:circularity}). What remains are two labour-market series moving in the same direction: a hiring gradient that turned against more exposed occupations after 2022 and a pay gap that has been compressing since 2017, neither of which this design attributes to generative AI. The fixed effects absorb shocks common to a broad occupational group or a region, so those forces would have to have hit more exposed occupations differentially within the same major group and region to account for the pattern. Remote work is exactly such a shock, and it is not clearly separable from AI exposure here.

\section{Validity}
\label{s:validity}

Having established how the index behaves as an instrument, this section asks whether it predicts reported or observed AI use, adds information beyond existing exposure measures, and travels to an independent survey.

\subsection{Predicting Reported AI Use}
\label{ss:predict_aiuse}

The SES asks separately about the use of four kinds of workplace technology: automated machinery, software that simplifies complex mental tasks, information-sharing technologies, and software described as having artificial intelligence. Only the last item names AI, and the four differ in how closely they should track an index of exposure to language models. Three are worker-facing software tools and are likely related to AI exposure; automated machinery refers to physical, embodied digital technology and should not. That gives convergent and discriminant predictions from the same battery. \Cref{tab:digitaltools} reports. Every column conditions on 1-digit SOC major group to limit the comparison to differences between workers in the same broad occupational class.

The pattern is the expected one. Exposure predicts use of all three worker-facing software tools, most strongly information-sharing technologies, and is unrelated to automated machinery: the point estimate there is negative and cannot be distinguished from zero. These differences are conditional on occupational class, so they are not simply a reflection of the fact that professional workers are more exposed to AI and more likely to use software than manual workers.

\textcite{lambert2026broken} argue that measures of AI exposure do not isolate susceptibility to AI, being bound up with other software and remote work more generally. The evidence here is consistent with that concern. Conditioning reported AI use on the worker's use of the other three technologies reduces the exposure gradient from 0.226 to 0.149. In this descriptive decomposition, roughly one-third of the unadjusted association is absorbed by measured digital tooling, while roughly two-thirds remains conditional on it. The pattern is stable across the panel: the unadjusted marginal effect ranges from 0.186 to 0.250 across raters, and the adjusted effect from 0.128 to 0.161; every estimate remains significant (\cref{tab:appB:digitaltools_raters}). GAISI therefore reflects a wider digital-work gradient but retains predictive content not absorbed by the measured, correlated digital tools.

\begin{table}[htbp]
\centering
\caption{Exposure and the Use of Workplace Technologies} \label{tab:digitaltools}
\begin{threeparttable}
\begin{tabularx}{\textwidth}{l *{5}{>{\centering\arraybackslash}X}}
\toprule
& \textbf{(1)} & \textbf{(2)} & \textbf{(3)} & \textbf{(4)} & \textbf{(5)} \\
& \textit{Automated} & \textit{Decision-} & \textit{Info-} & \textit{AI} & \textit{AI software,} \\
& \textit{machinery} & \textit{support sw} & \textit{sharing} & \textit{software} & \textit{adjusted} \\
\midrule
Exposure (per IQR) & $-0.037$ & 0.178 & 0.239 & 0.226 & 0.149 \\
                   & (0.036)  & (0.035) & (0.038) & (0.034) & (0.026) \\
$p$                & 0.316    & $<$0.001 & $<$0.001 & $<$0.001 & $<$0.001 \\
\addlinespace
\textit{of which:} within-rater & (0.022) & (0.024) & (0.027) & (0.027) & (0.024) \\
\phantom{of which:} between-rater & (0.027) & (0.023) & (0.026) & (0.020) & (0.011) \\
Rater share of variance & 0.62 & 0.50 & 0.51 & 0.38 & 0.19 \\
\bottomrule
\end{tabularx}
\begin{tablenotes}[flushleft]
\footnotesize
\item \textit{Note:} Average marginal effects from logistic regressions of each technology-use item on generative AI exposure, in a sample of 5,762 workers. Exposure enters as a percentile rank. The reported coefficients are the change in the probability of using that technology associated with a one interquartile range increase in a worker's exposure relative to other workers. All columns control for total task load, survey year and SOC major group; column (5) additionally controls for the worker's use of the three technologies in columns (1)--(3). Each column is estimated once per rater and the nine estimates combined by the Rubin-style approach, so the reported standard error contains both the sampling variation within a rater and the disagreement between them; the two components are shown separately, and the final row gives the between-rater share of the Rubin-combined variance. Sampling standard errors are clustered by 4-digit occupation. \Cref{tab:appB:digitaltools_raters} reports each rater separately.
\item \textit{Source:} SES~2023--24, ages 20--65. Authors' calculations.
\end{tablenotes}
\end{threeparttable}
\end{table}

\subsection{Value Added over Existing Measures}
\label{ss:validity_existing}
A central question for any new exposure index is what it contributes beyond existing measures. Its rank correlations with established occupation-level technology exposure measures line up with how close each measure's target construct sits to generative AI (\cref{tab:appB:ses_gaisi_corr}). Measures built for earlier waves of automation correlate negatively or not at all: $-0.58$ with \textcite{Frey2017TheComputerisation}'s automation probability and $-0.47$ with \textcite{Webb2020TheMarket}'s robot exposure, rising to 0.21 for suitability for machine learning \parencite{Brynjolfsson2018WhatEconomy}, which targets a predictive form of AI. Measures built for generative AI correlate at 0.75 to 0.80. 

The ordering also helps to examine whether these correlations are an artefact of shared machine authorship. \textcite{Eloundou2024GPTsLLMs} publish two versions of their index, one rated by human annotators and one by GPT-4. Exposure correlates with the human-rated version at 0.758 and with the machine-rated version at 0.753, and the intervals across raters overlap almost entirely. In other words, agreement does not depend on the benchmark having been produced by a language model, which is not the pattern one would expect if the index captured only shared bias in how LLMs interpret task descriptions.

Correlations of that size also raise the opposite question: whether the index adds anything beyond the existing measures. This is a question about prediction. \Cref{tab:ame_ai_adoption} enters the existing measures alongside GAISI in blocks, ordered by construct distance. The pre-generative-AI measures barely move the exposure coefficient, from 0.226 to 0.220: indices built for earlier waves of automation say little about who uses AI at work that GAISI does not capture. By contrast, the human-derived generative-AI measures attenuate the estimate to 0.160, which is expected from measures of the same construct. Adding the GPT-4-rated version of EMMR on top of them changes almost nothing further, at 0.154 --- the LLM-generated benchmark contributes little that the human-rated one has not already contributed. Around two-thirds of the association remains conditional on the full set of exposure measures, and it does so under every rater in the panel.

The between-rater share of the remaining uncertainty behaves differently here than in \cref{tab:digitaltools}. It rises as the blocks are added, from 0.38 to 0.53, where conditioning on the worker's own tools had lowered it. The benchmarks are themselves exposure measures: they absorb the part of the index that the raters agree about, and leave behind the part where they do not.

\begin{table}[htbp]
\centering
\begin{threeparttable}
\caption{\TabNineCaption}
\label{tab:ame_ai_adoption}
\setlength{\tabcolsep}{6pt}
\begin{tabularx}{\textwidth}{l *{4}{>{\centering\arraybackslash}X}}
\toprule
& \textbf{(1)} & \textbf{(2)} & \textbf{(3)} & \textbf{(4)} \\
& \textit{Exposure} & \textit{+ pre-GenAI} & \textit{+ GenAI,} & \textit{+ GenAI,} \\
& \textit{alone}    & \textit{measures}    & \textit{human-rated} & \textit{LLM-rated} \\
\midrule
Exposure (per IQR)  & 0.226 & 0.220 & 0.160 & 0.154 \\
                    & (0.034) & (0.036) & (0.038) & (0.036) \\
$p$                 & $<$0.001 & $<$0.001 & $<$0.001 & $<$0.001 \\
\addlinespace
\textit{of which:} within-rater   & (0.027) & (0.027) & (0.025) & (0.025) \\
\phantom{of which:} between-rater & (0.020) & (0.022) & (0.027) & (0.025) \\
Rater share of variance & 0.38 & 0.44 & 0.56 & 0.53 \\
\addlinespace
Range across raters & 0.186--0.250 & 0.175--0.249 & 0.108--0.195 & 0.107--0.186 \\
\midrule
Benchmark measures added & --- & 5 & 8 & 9 \\
Observations & 5{,}683 & 5{,}683 & 5{,}683 & 5{,}683 \\
\bottomrule
\end{tabularx}
\begin{tablenotes}[flushleft]
\footnotesize
\item \textit{Note:} Average marginal effect of generative AI exposure on the probability that a worker reports using software with artificial intelligence, as competing occupational exposure measures are added in blocks. Column (1) enters exposure alone; column (2) adds the five measures built for pre-generative-AI technologies; column (3) adds the three generative-AI measures derived from human ratings or crowdsourced capability links; column (4) adds the GPT-4-rated version of EMMR, the only benchmark generated by a language model. Blocks follow the construct ordering in \cref{tab:appB:ses_gaisi_corr}. All measures enter as percentile ranks. Every column controls for workers' total task load, survey year, 1-digit SOC major group, sex, age, ethnicity, region and education; column (1) is therefore the same regression as column (4) of \cref{tab:digitaltools}. Each column is estimated once per rater and pooled by the Rubin-style approach; the reported standard error combines sampling variation within a rater with disagreement between raters, and the two components are shown separately. Sampling standard errors are clustered by 4-digit occupation.
\item \textit{Source:} SES~2023--24, ages 20--65. Authors' calculations.
\end{tablenotes}
\end{threeparttable}
\end{table}

\subsection{Agreement with Observed AI-Assistant Use}
\label{ss:aei_traffic}

The two preceding subsections compare the worker-level index with outcomes the same workers report or with indices built by other researchers. This subsection moves to the task-level ratings and compares them with a record of what people ask an AI assistant to do. Anthropic's Economic Index publishes 630 conversation clusters built from use of the Claude assistant \parencite{handa2025economictasksperformedai}; each cluster describes a related use case (e.g., ``Create educational research content and teaching materials'', ``Draft professional emails about technical or business matters'') and carries the share of all conversations it accounts for. Mapping the clusters onto the 44 SES tasks gives every task a share of assistant traffic. \Textcite{NBERw35110} argue that exposure should be anchored in observed usage. The traffic shares supply an external benchmark, which is neither reported by workers in SES nor another researcher's exposure index.

The units of observation are the 1{,}100 task-occupation cells. Traffic shares are regressed on exposure. This analysis says nothing about which jobs or workers use an assistant.  Every specification includes occupation fixed effects, so the comparison runs within occupation: among the SES tasks, are the ones the panel rates as more exposed also the ones people bring to an assistant? Both variables enter as percentile ranks, so a coefficient is the movement in traffic rank per interquartile range of exposure rank, in line with the rest of the paper and avoiding a handful of high-traffic tasks being overly influential. Each regression is run once per rater and the nine estimates pooled by our Rubin-style approach, so the standard error carries both sampling variation within a rater and disagreement between raters in the multiverse.

A task one interquartile range higher in exposure sits 0.511 interquartile ranges higher in assistant traffic (SE 0.108, $p<0.001$; \cref{tab:aeitraffic}, first row). Rater-specific estimates run from 0.469 to 0.564, and all nine are significant. The between-rater share of the Rubin-combined variance accounts for 0.09 of the variance of the pooled estimate. The ordering the ratings impose on tasks agrees with where assistant traffic actually goes, consistently under every rater in the panel.

That agreement does not depend on what model mapped the clusters in this comparison. Compared with a Gemini-generated mapping, re-mapping the 630 clusters with DeepSeek-V4-Pro, a model not in the rater panel, raises the pooled coefficient to 0.577 (SE 0.104). A mapping built instead from a sentence-embedding model, which performs no generative reasoning, still returns a coefficient of 0.388 (SE 0.150, $p=0.010$). Raters disagree more under that mapping, a between-rater share of 0.25, and estimates spread from 0.302 to 0.526, but all nine remain significant.

The traffic evidence refers to overall exposure. Replacing the composite with the direct-exposure component E1 alone gives a coefficient of 0.468 under the Gemini mapping, close to the composite's 0.511. A chatbot (E1)-specific reading of these traffic shares is not supported by the analysis: assistant conversations concentrate in tasks the panel rates as exposed, without separating direct chatbot suitability from exposure that runs through software integration. While insightful, one needs to bear in mind that the benchmark covers one assistant's users rather than the population of AI users.

\begin{table}[htbp]
\centering
\begin{threeparttable}
\caption{Assistant Conversation Share and Task-Level Exposure}
\label{tab:aeitraffic}
\begin{tabularx}{\textwidth}{l *{5}{>{\centering\arraybackslash}X}}
\toprule
 & \textbf{Coef.} & \textbf{SE} & \textbf{$p$} & \textbf{Range across} & \textbf{Rater share} \\
\textbf{Cluster-to-task mapping} & \textbf{(per IQR)} & & & \textbf{9 raters} & \textbf{of variance} \\
\midrule
Gemini~2.5~Flash    & 0.511 & (0.108) & $<$0.001 & 0.469--0.564 & 0.09 \\
DeepSeek-V4-Pro     & 0.577 & (0.104) & $<$0.001 & 0.493--0.633 & 0.21 \\
Embedding           & 0.388 & (0.150) & 0.010    & 0.302--0.526 & 0.25 \\
\bottomrule
\end{tabularx}
\begin{tablenotes}[flushleft]
\footnotesize
\item \textit{Note:} The unit of observation is the task-occupation cell (1,100 cells); the dependent variable is the share of all AI-assistant conversations accounted for by the task, and the regressor is GAISI exposure for that cell. Both enter as percentile ranks, so the coefficient is the movement in conversation-share rank associated with a one interquartile range increase in the exposure rank, within occupation: all specifications include occupation fixed effects, and standard errors are clustered on task (44 clusters). Each row is estimated once per rater and the nine estimates combined by the Rubin-style approach, so the reported standard error contains both the sampling variation within a rater and the disagreement between raters; the final column gives the between-rater share of the Rubin-combined variance. Rows differ only in how the 630 conversation clusters were mapped onto the 44 SES task handles; every rater-specific estimate is positive and significant at the 5 per cent level in all three rows. DeepSeek-V4-Pro is from a vendor not represented among the nine raters.
\item \textit{Source:} Anthropic Economic Index conversation clusters \parencite{handa2025economictasksperformedai} and SES~2023--24 task handles. Authors' calculations.
\end{tablenotes}
\end{threeparttable}
\end{table}

\subsection{Calibration on Observed Diffusion}
\label{ss:vintage}
The nine raters span model generations released between 2024 and 2026, and the later ones may have been trained on text describing the diffusion of the technology they are asked to assess. Conditioning on what an earlier generation said isolates change relative to that baseline, though it cannot distinguish corpus feedback from a better reading of the occupation-task information. Nonetheless, it distinguishes two concerns when comparing ratings from multiple generations: whether exposure levels drift upward, and whether ratings drift towards the particular tasks people use AI for.

There is no general upward drift in levels. Within Google's lineage mean exposure falls across generations, from 0.40 for Gemini~1.5~Pro to 0.38 and then 0.29 for Gemini~3.1~Pro, the lowest of the nine raters; OpenAI's rises from 0.35 to 0.46 and returns to 0.35 for its most recent model. Later raters are not more generous about how much of a job an assistant could take on.

We tested whether observed activity predicts a later model's rating of a task-occupation cell once an earlier model's rating of the same cell is held fixed, comparing each vendor's generations in order against the reverse ordering, where no such feedback loop is possible. At the job level, with the worker's own reported AI use as the measure of activity and occupation fixed effects throughout, the average drift across the six lineage pairs is $-0.003$ of an interquartile range, and none of the six pair-specific drift estimates is statistically significant. Reported use is measured with error and the residual variation it has to predict is small---adjacent generations correlate at about 0.9---so this is not a precisely estimated zero.

At the task level, where activity is measured by the share of assistant conversations a task accounts for, one pair shows statistically significant drift. Between Gemini~1.5~Pro and Gemini~3.1~Pro, a one-IQR increase in assistant traffic is associated with a 0.170-IQR upward drift in exposure ratings (Bonferroni-adjusted \(p=0.039\) across twelve comparisons). The remaining five lineage pairs show no statistically significant drift. Something in that lineage step moved the ratings towards the tasks people had already brought to an assistant. The design cannot say whether the corpus or a better reading of the work is responsible.

\subsection{Circularity and the Limits of Validation}
\label{ss:circularity}

Circularity is a possible limitation of validation against worker-reported AI use. If the raters have learned the same public account of which jobs are exposed, and if workers draw on that account when deciding whether the software they use counts as AI, their common response to it could yield $cov(b,e_2)>0$ in the notation of \cref{eq:eq4}. Neither condition nor the resulting covariance is observed directly. The panel cannot detect or average away any such $b$, because, if present, it is shared across raters. Shared framing would create a source of upward bias in the association between GAISI and reported AI use. Other measurement errors could attenuate the same association, so the direction and magnitude of the net bias are not identified. We treat the reported association as descriptive evidence of predictive content, not as an unbiased estimate of the relationship between latent exposure and actual use.

The horse race in \cref{tab:ame_ai_adoption} is the most informative check available. GAISI remains associated with reported AI use after nine existing exposure measures enter together, including human- and LLM-rated generative-AI indices; its marginal effect falls from 0.226 to 0.154. This shows that GAISI contains predictive information not spanned by those measures. It does not rule out circularity, because the outcome remains a worker report and the measures may draw on overlapping accounts of AI exposure. The association with decision-support software (\cref{tab:digitaltools}) provides discriminant-validity evidence, but the difference between the two coefficients cannot identify a framing effect: the questions name different technologies and use different language.

The labour-market estimates do not resolve the issue. Pay and vacancies are unlikely to pass through a shared public narrative of AI exposure, but neither design identifies a distinct effect of AI exposure once the relevant competing explanations are considered. Assistant traffic supplies a behavioural benchmark at the task level (\cref{ss:aei_traffic}), although later models may have learned from the same diffusion that the traffic records (\cref{ss:vintage}). The evidence therefore supports GAISI's incremental predictive content while leaving the existence and possible size of a shared component $b$ unidentified. 

\subsection{Portability and Consequences of Use}
\label{ss:generalisation}
The same classification procedure applied to the OECD Survey of Adult Skills (PIAAC) reproduces the UK ordering. PIAAC uses different task items and response scales and covers a slightly higher share of working hours (38\%). Mean exposure (0.49) and median (0.54) exceed the SES benchmarks, as the level comparisons in \cref{ss:rating_precision} would lead one to expect, but dispersion is similar (SD = 0.13, 80--20 percentile range = 0.17) and the correlation with SES-GAISI across broad occupation-industry cells reaches $\rho = 0.94$ $(p < 0.001)$. Again the ordering transfers better than the levels.

Splitting the cells finer---crossing occupation and industry with education, sex and age, from 32 cells to 246---moves the weighted correlation only from 0.940 to 0.911. Carrying the UK ranking down to individual PIAAC workers preserves it as well. Across 84,560 workers in 22 of the 30 participating countries whose public-use files report occupation, a worker's own PIAAC exposure rank rises by 0.73 of an interquartile range (SE 0.06) per interquartile range of the ranking imported from the UK, conditional on country, sex, education, age, and that worker's measured literacy and numeracy. Without those controls, the coefficient is 0.80. In other words, the agreement survives conditioning on individuals' schooling and skill. The imported ranking accounts for 51\% of the variation the controls leave unexplained, and about three-quarters of the 70\% accounted for by a full set of cell dummies, which is substantial for any cell-level measure (\cref{tab:appB:piaac_port}). The UK ranking enters as a cell aggregate and carries no within-cell variation, so this is not an individual-level replication of $\rho = 0.94$; what it adds is the covariate set.

Shared vendor does not explain the agreement. Gemini~2.5~Pro produced the PIAAC scores and also sits in the nine-rater panel, so the two surveys might agree because one model agrees with itself. It is the weakest of the nine predictors (0.706, against 0.748 for the strongest), and the three Gemini raters as a group (0.756) do no better than the six others (0.767).

Using the index does not import a demographic gradient. Rater prompts omitted worker characteristics, but LLMs might still associate occupations or tasks with demographic groups, and a check run on one rater cannot see a gradient peculiar to another. Regressing residual exposure, after accounting for the worker's task profile and 2-digit occupation, on sex, age, ethnicity and education returns uniformly small associations under every rater in the panel: no coefficient exceeds 0.015 of an interquartile range of the exposure rank, and $R^2$ ranges from 0.4\% to 1.0\%, below 1\% for eight of the nine. At the ensemble average $R^2 = 0.8\%$ (\cref{tab:appB:gaisi_bias}).

\section{Conclusion}
\label{s:conclusion}

This paper develops GAISI, a task-based measure of UK jobs' exposure to generative AI, using nine large language models to rate 1,100 task--occupation cells. The ratings support a clearer ordering of jobs than
a common exposure level. Pairwise rank correlations across raters range from 0.74 to 0.92. Level statistics are less stable: depending on the model, the share of British jobs scoring above 0.5 ranges from under
0.1\% to 38\%, and mean exposure ranges from 0.29 to 0.46. A measure constructed with one model would conceal this dependence on the choice of rater.

This distinction matters for the use of LLMs as measurement instruments. A nine-rater ensemble treats the model as an analytic choice: each downstream analysis is estimated under every rater, with the rater-specific results reported alongside a pooled summary that incorporates their dispersion. Prompt sensitivity is assessed separately. Because the panel is purposive rather than sampled from a defined population of models, its spread describes sensitivity across the raters considered here; it is not a calibrated estimate of all possible measurement error. The same reporting requirements apply to public classifications such as the UK Standard Skills Classification, not only to academic exposure indices.

The external evidence gives reasons to take the resulting ordering seriously, while leaving relevant questions unresolved. GAISI's association with worker-reported AI use falls from 0.226 to 0.154 after nine competing
exposure measures enter together, including human-rated generative-AI indices, but remains positive under every rater. At the task level, exposure rank is associated with observed assistant traffic under both generative and embedding-based mappings of conversation clusters. Applying the procedure to PIAAC, a different task survey, also produces a similar ordering across occupation--industry cells. These results establish conditional associations and portability across measurement settings. They do not establish whether the ratings contain a biased component shared through overlapping training corpora, or its size if they do. In particular, common discourse may enter both GAISI and workers' reports of AI use, so the worker-level association should not be read as an unbiased relationship between latent exposure and actual use.

The labour-market applications warrant the same restraint. Mean worker exposure rank rose by 4.5 percentile points between 2017 and 2023/24, almost entirely because employment shifted between occupations rather than because task profiles changed within them. Relative pay and job adverts also moved against more exposed occupations after 2022. Both series, however, were already moving before ChatGPT, and the vacancy association disappears once teleworkability is allowed its own post-2022 path. Under that specification,
seven raters produce significant positive coefficients and two produce small negative coefficients, while the pooled estimate is near zero. The paper therefore does not identify an effect of generative AI on pay or labour demand. It documents labour-market patterns associated with the exposure ranking and shows how their interpretation changes when purposive rater choice and competing occupational trends are made visible.

GAISI has practical limits. The SES task battery covers about one-third of working time. Treating all unobserved work as unexposed lowers discrimination from 0.72 to 0.66 under every rater. The panel cannot detect measurement bias if shared across raters. Ratings may also become dated as models and patterns of use change. We find no general upward drift in exposure levels across model generations, but one of six lineage comparisons shows task-specific movement towards observed assistant use. The measure therefore requires periodic rerating and continued comparison across models. Its broader contribution is a procedure for using LLM judgements in empirical work while exposing model choice, separating stable rankings from unstable levels, and carrying observed disagreement into the resulting estimates.

\printbibliography


\clearpage
\begin{appendices}
\appendixpage
\addappheadtotoc

\counterwithin{figure}{section}
\counterwithin{table}{section}
\crefalias{section}{appendix}

\section{Appendix: Rater Prompt and Task Battery}
\label{app:A:prompt}

\begin{lstlisting}[style=prompt]
Your goal is to assess the probability of different levels of AI exposure for each task within a specific occupation category.

Before we begin, please review the following occupation details:
<occupation_details>
{occupation_details}
</occupation_details>

You will analyze a category of related tasks for Occupation Code {occupation_code}. For each task, determine the probability of different AI exposure levels.
<category_name>
{category_name}
</category_name>

You will analyze the following list of tasks:
<task_list>
{task_list}
</task_list>

Instructions:
1. Analyze each task in the context of the occupation details provided.
2. Consider how an average worker in this occupation would typically perform each task using
   existing tools.
3. Assess how LLMs could potentially assist with each task, focusing on new capabilities
   beyond existing tools.
4. For EACH TASK, calculate the probability of it falling into each of the following AI
   exposure levels:
   - E0: No Exposure (LLMs do not meaningfully reduce time by 25% or more)
   - E1: Direct Exposure (LLMs alone can reduce time by at least 25%)
   - E2: Exposure via Imaginable LLM-Powered Applications (LLMs + additional software could
         reduce time by at least 25%)
   - E3: Exposure given Image Capabilities (Image processing significantly reduces time by
         25% or more)
5. Provide a brief justification for each probability distribution.

In your analysis for each task, include:
a. Task summary in the context of the occupation
b. Existing tools typically used for this task
c. Potential LLM capabilities that could assist with the task
d. Arguments for and against each exposure level

First provide a high-level analysis of the entire category in this occupation context. Then analyze each task individually, giving probability distributions that sum to 1.0 (expressed as decimals).

Output Format:
Provide your response in the following JSON format:
```json
[
  {
    "task_handle": "task1_handle",
    "occupation_code": "{occupation_code}",
    "task_description": "Description of task 1",
    "probabilities": {
      "E0": 0.XX,
      "E1": 0.XX,
      "E2": 0.XX,
      "E3": 0.XX
    },
    "justification": "One sentence justification for probabilities"
  },
  ...
]

Remember to consider the incremental impact of LLMs beyond pre-existing tools and technologies, and focus on the potential 25% or more time reduction for each exposure level.
\end{lstlisting}

\clearpage


\subsection*{Task Battery}
\begin{longtable}{P{0.22\textwidth} c P{0.60\textwidth}}
\caption{\TabOneCaption}\label{tab:taskcats}\\
\toprule
\textbf{Task Category} & \textbf{N} & \textbf{Constituent SES Tasks} \\
\midrule
\endfirsthead
\toprule
\textbf{Task Category} & \textbf{N} & \textbf{Constituent SES Tasks} \\
\midrule
\endhead

\multicolumn{3}{l}{\textit{Manual}} \\[0.2em]
Manual & 4 & Carrying, pushing or pulling heavy objects; working for long periods on physical activities; mending, repairing, assembling, constructing or adjusting things; knowledge of how to use or operate tools, equipment or machinery \\[0.4em]

\multicolumn{3}{l}{\textit{Cognitive}} \\[0.2em]
Reading & 3 & Reading written information such as forms, notices or signs; short reports, letters or emails; long reports, manuals, articles or books \\[0.4em]
Writing & 3 & Writing forms or notices; short documents; long documents with correct spelling and grammar \\[0.4em]
Numeracy & 3 & Basic arithmetic; decimals, percentages or fractions; advanced mathematical or statistical procedures \\[0.4em]
Planning \& Organising & 5 & Planning own activities; planning others' activities; organising own time; thinking ahead; developing plans \\[0.4em]
Expertise \& Innovation & 3 & Specialist knowledge; keeping up-to-date with developments; developing improved processes, products or services \\[0.4em]
Problem Analysis & 4 & Spotting problems; diagnosing causes; proposing solutions; analysing complex problems \\[0.4em]

\multicolumn{3}{l}{\textit{Interpersonal / Management}} \\[0.2em]
Professional Communication & 3 & Instructing or training; making presentations; persuading or influencing \\[0.4em]
Client Interaction & 4 & Dealing with people; counselling, advising or caring; selling; product or service knowledge \\[0.4em]
Collaboration & 3 & Teamworking; listening carefully; cooperating \\[0.4em]
Emotion \& Impression & 4 & Managing own feelings; handling others' feelings; looking or sounding the part \\[0.4em]
Management & 5 & Motivating staff; controlling resources; coaching; developing careers; strategic decisions \\

\bottomrule
\end{longtable}

\begin{table}[htbp]
\centering
\caption{\TabTwoCaption} \label{tab:taskimportance}
\begin{threeparttable}
\begin{tabularx}{\textwidth}{l *{5}{>{\centering\arraybackslash}X}}
\toprule
\textbf{Task Categories} & \textbf{N} & \textbf{Mean} & \textbf{SD} & \textbf{10th pct} & \textbf{90th pct} \\
\midrule
Reading                        & 5,783 & 0.71 & (0.26) & 0.33 & 1.00 \\
Writing                        & 5,784 & 0.60 & (0.31) & 0.17 & 1.00 \\
Numeracy                       & 5,784 & 0.49 & (0.33) & 0.00 & 1.00 \\
Planning and Organising        & 5,784 & 0.67 & (0.23) & 0.35 & 0.95 \\
Expertise and Innovation       & 5,784 & 0.71 & (0.22) & 0.42 & 1.00 \\
Problem Analysis               & 5,782 & 0.73 & (0.25) & 0.38 & 1.00 \\
Professional Communication     & 5,783 & 0.52 & (0.28) & 0.17 & 0.92 \\
Client Interaction             & 5,784 & 0.62 & (0.22) & 0.31 & 0.94 \\
Collaboration                  & 5,784 & 0.80 & (0.24) & 0.50 & 1.00 \\
Emotion and Impression         & 5,784 & 0.68 & (0.21) & 0.38 & 0.94 \\
Management                     & 5,783 & 0.26 & (0.36) & 0.00 & 0.85 \\
\midrule
\textbf{Total Task Load (Importance-weighted)}
                               & 5,784 & 25.88 & (7.06) & 16.50 & 35.00 \\
\bottomrule
\end{tabularx}
\begin{tablenotes}[flushleft]
\footnotesize
\item \textit{Note:} Worker-reported importance of each task category, scored 1 for an essential task down to 0 for one that is not at all important, and averaged over the tasks in the category. The final row is the sum over all 44 tasks, so it is the number of essential-equivalent tasks a job comprises and is the denominator in \cref{eq:eq2}. Weighted for survey design and non-response.
\item \textit{Source:} SES~2023--24 (UK). Authors' calculations.
\end{tablenotes}
\end{threeparttable}
\end{table}

\subsection{Grounding in the Raters' Written Justifications}
\label{ss:appA:grounding}

Alongside every score, each rater wrote a short justification. DeepSeek~V4~Flash, which is not in the rater panel, coded 18,583 justifications against fourteen fixed tags: five affordance cues, human constraints, explicit limitations, integration, and discourse markers. In \cref{tab:appA:grounding}, affordance cues predict higher scores and human constraints and explicit limitations predict lower ones. The relationships remain within task--occupation cells, where identification comes from differences between raters' readings of the same cell.

The result is not specific to the coding model. Gemini~2.5~Flash recoded 10,989 justifications; its affordance, human-constraint and limitation counts correlate 0.75, 0.85 and 0.87 with the DeepSeek counts and have near-identical means. Across all fourteen tags, the coders agree 87.4 per cent of the time on average. \Cref{tab:appA:grounding_raters} reports the estimates separately by rater.

\begin{table}[htbp]
\centering
\begin{threeparttable}
\caption{Exposure Scores and Coded Justification Cues}
\label{tab:appA:grounding}
\begin{tabularx}{\textwidth}{l *{2}{>{\centering\arraybackslash}X}}
\toprule
 & \textbf{(1)} & \textbf{(2)} \\
\midrule
Affordance cue count                              & 0.196    & 0.090    \\
                                                  & (0.006)  & (0.004)  \\
Affordance cue count, squared                     & $-$0.048 & $-$0.021 \\
                                                  & (0.002)  & (0.001)  \\
Human-constraint cue count                        & $-$0.064 & $-$0.019 \\
                                                  & (0.003)  & (0.002)  \\
Explicit-limitation cue count                     & $-$0.106 & $-$0.056 \\
                                                  & (0.003)  & (0.002)  \\
\addlinespace
Average marginal effect, affordance count         & 0.084    & 0.040    \\
\midrule
Model $\times$ run fixed effects                  & Yes      & Yes      \\
Task-occupation fixed effects                     & No       & Yes      \\
Discourse, integration and length controls        & Yes      & Yes      \\
Observations                                      & 18{,}583 & 18{,}583 \\
$R^2$                                             & 0.574    & 0.856    \\
\bottomrule
\end{tabularx}
\begin{tablenotes}[flushleft]
\footnotesize
\item \textit{Note:} The unit is a justification written by one rater for one task--occupation cell, and the outcome is its accompanying 0--1 exposure score. DeepSeek~V4~Flash assigned all fourteen tags. Both columns control for integration, hedging, contrast and a cubic polynomial in justification length; column~(1) absorbs model~$\times$~run and column~(2) also absorbs task--occupation. Standard errors in parentheses are clustered on task--occupation cell; all displayed coefficients have $p<0.001$. A total of 117 justifications were missing or not encoded because of server error.
\item \textit{Source:} Rater justifications for the 44 SES task handles $\times$ 25 SOC~2010 occupation groups. Authors' calculations.
\end{tablenotes}
\end{threeparttable}
\end{table}

\begin{table}[htbp]
\centering
\begin{threeparttable}
\caption{Exposure Scores and Coded Justification Cues, by Rater}
\label{tab:appA:grounding_raters}
\setlength{\tabcolsep}{5pt}

\begin{tabularx}{\textwidth}{l *{5}{>{\centering\arraybackslash}X}}
\toprule
& \textbf{Affordance} & & \textbf{Standardised} & \textbf{Within-cell} & \\
\textbf{Rater} & \textbf{count} & \textbf{AME} & \textbf{effect} & \textbf{coefficient} & \textbf{$N$} \\
\midrule
Gemini~3.1~Pro & 0.308 \: (0.033) & 0.191 & 0.607 & 0.047 \: (0.004) & 1{,}080 \\
Gemini~2.5~Pro & 0.246 \: (0.031) & 0.133 & 0.440 & 0.089 \: (0.004) & 1{,}078 \\
GLM~5.2        & 0.241 \: (0.030) & 0.104 & 0.335 & 0.030 \: (0.004) & 1{,}100 \\
Gemini~1.5~Pro & 0.236 \: (0.030) & 0.103 & 0.363 & 0.044 \: (0.003) & 5{,}489 \\
GPT-5.5        & 0.128 \: (0.033) & 0.044 & 0.170 & 0.022 \: (0.003) & 1{,}094 \\
GPT-5.6-Luna   & 0.126 \: (0.026) & 0.050 & 0.212 & 0.009 \: (0.003) & 1{,}062 \\
GPT-4o         & 0.124 \: (0.016) & 0.052 & 0.214 & 0.004 \: (0.002) & 5{,}500 \\
GPT-5.1        & 0.100 \: (0.019) & 0.031 & 0.139 & 0.013 \: (0.004) & 1{,}080 \\
GPT-5.4        & 0.081 \: (0.020) & 0.020 & 0.079 & 0.015 \: (0.003) & 1{,}100 \\
\bottomrule
\end{tabularx}

\begin{tablenotes}[flushleft]\footnotesize
\item \textit{Note:} Rater-specific estimates corresponding to \cref{tab:appA:grounding}. The first columns report the linear affordance-count coefficient, its average marginal effect accounting for the squared term, and a standardised effect. ``Within-cell coefficient'' adds task--occupation fixed effects. DeepSeek~V4~Flash supplies all cue codes. All coefficients have $p<0.05$; standard errors in parentheses are clustered on task--occupation cell.
\item \textit{Source:} SES~2023--24 task handles $\times$ SOC~2010 occupation groups. Authors' calculations.
\end{tablenotes}

\end{threeparttable}
\end{table}


\section{Appendix: Properties and Validity}
\label{app:B:eval}

The tables below run in the order of the body: those supporting \cref{s:properties}, on how the index behaves as an instrument, come first, followed by those supporting \cref{s:validity}.

\begin{table}[htbp]\centering
\begin{threeparttable}
\caption{Rater-Panel Variance Decomposition, Ensemble Reliability and Cross-Rater Rank Agreement}
\label{tab:appB:raterpanel}
\setlength{\tabcolsep}{6pt}

\begin{tabularx}{\textwidth}{l *{2}{>{\centering\arraybackslash}X}}
\toprule
\multicolumn{3}{l}{\textit{Panel A. Variance components of the exposure rating}} \\
\midrule
 & \textbf{Variance} & \textbf{\% of rater variance} \\
Task--occupation cell (signal $+$ shared bias) & 0.0376 & --- \\
Vendor house style             & 0.0014 & 11.2 \\
Model version within vendor    & 0.0021 & 16.8 \\
Cell $\times$ rater (idiosyncratic) & 0.0090 & 72.0 \\
\addlinespace
Share of total variance between cells & \multicolumn{2}{c}{75.0\%} \\
\midrule
\multicolumn{3}{l}{\textit{Panel B. Reliability of a $k$-rater ensemble}} \\
\midrule
$k$ & 1 \quad 2 \quad 3 \quad 4 \quad 5 & 6 \quad 7 \quad 8 \quad 9 \\
$\rho_k$ & 0.750 \enspace 0.857 \enspace 0.900 \enspace 0.923 \enspace 0.937 & 0.947 \enspace 0.954 \enspace 0.960 \enspace 0.964 \\
\midrule
\multicolumn{3}{l}{\textit{Panel C. Pairwise rank correlation between raters (36 pairs)}} \\
\midrule
 & \textbf{Min \quad 25th \quad Median} & \textbf{Mean \quad 75th \quad Max} \\
Spearman $\rho$ & 0.743 \enspace 0.780 \enspace 0.807 & 0.813 \enspace 0.848 \enspace 0.915 \\
\bottomrule
\end{tabularx}

\begin{tablenotes}[flushleft]\footnotesize
\item \textit{Note:} All panels use the nine raters applied to the same 1,100 task--occupation cells. \textit{Panel A} decomposes ratings after repeat runs are averaged. Cell variance contains $G^*$ and any shared bias $b$; vendor and version capture the systematic rater component $\delta$, and cell $\times$ rater captures $u$ (\cref{eq:eq4}). Percentages are shares of rater variance. The 75.0\% between-cell share is single-rating reliability when level differences between raters count as error, unlike the consistency coefficients in \cref{tab:raterreliability}. \textit{Panel B} applies the Spearman--Brown formula, assuming interchangeable, equally reliable raters; averaging shrinks $\delta$ and $u$ but not $b$. \textit{Panel C} summarises the 36 pairwise Spearman correlations across the 1,098 cells rated by all nine.
\end{tablenotes}

\end{threeparttable}
\end{table}

\begin{table}[htbp]
\centering
\begin{threeparttable}
\caption{Exposure under Prompt Permutation}
\label{tab:appB:prompt_permutation}
\setlength{\tabcolsep}{4pt}
\begin{tabularx}{\textwidth}{l *{6}{>{\centering\arraybackslash}X}}
\toprule
 & \textbf{Rank agr.} & \textbf{Rank agr.} & \textbf{AUC} & & & \\
\textbf{Prompt} & \textbf{same model} & \textbf{benchmark} & \textbf{AI use} & \textbf{$p$} & \textbf{Mean} & \textbf{SD} \\
\midrule
\multicolumn{7}{l}{\textit{Main prompt}} \\
Gemini~1.5~Pro (benchmark)      & ---   & 1.000 & 0.725 & ---   & 0.399 & 0.105 \\
Gemini~2.5~Pro                  & 1.000 & 0.796 & 0.714 & 0.033 & 0.382 & 0.130 \\
\addlinespace
\multicolumn{7}{l}{\textit{Prompt varied, Gemini~2.5~Pro throughout}} \\
Simplified wording              & 0.846 & 0.808 & 0.713 & 0.074 & 0.419 & 0.123 \\
\quad + no worked examples      & 0.816 & 0.789 & 0.715 & 0.127 & 0.411 & 0.103 \\
\quad + no existing-tools cue   & 0.802 & 0.781 & 0.706 & 0.094 & 0.434 & 0.094 \\
50\% time-saving threshold      & 0.796 & 0.773 & 0.722 & 0.659 & 0.393 & 0.099 \\
\addlinespace
\multicolumn{7}{l}{\textit{Rubric varied}} \\
Original EMMR rubric            & 0.569 & 0.573 & 0.685 & 0.004 & 0.536 & 0.094 \\
\bottomrule
\end{tabularx}
\begin{tablenotes}[flushleft]\footnotesize
\item \textit{Note:} Every row after the first two uses Gemini~2.5~Pro. ``Same model'' is Spearman rank agreement with that model at the main prompt; ``benchmark'' compares with Gemini~1.5~Pro. Both use the 1,096 cells observed in every variant. AUC measures discrimination of reported AI use among 5,740 workers; $p$ compares it with the benchmark AUC in a single survey-weighted \texttt{somersd} fit clustered on 4-digit occupation. Mean and SD are weighted worker-level exposure on the 0--1 scale; substantive estimates use ranks. The categorical EMMR rubric takes three values, so ties attenuate its rank agreement.
\item \textit{Source:} SES~2023--24, ages 20--65. Authors' calculations.
\end{tablenotes}
\end{threeparttable}
\end{table}

\begin{table}[htbp]
\centering
\begin{threeparttable}
\caption{The Weight on Latent Exposure Implied by Each Rater}
\label{tab:appB:e2_weight}
\setlength{\tabcolsep}{6pt}
\begin{tabularx}{\textwidth}{l *{5}{>{\centering\arraybackslash}X}}
\toprule
 & \textbf{Implied} & & & \textbf{AUC at} & \textbf{AUC gap} \\
\textbf{Rater} & \textbf{weight} & \textbf{SE} & \textbf{95\% CI} & \textbf{$\omega=0.5$} & \textbf{(derived $-$ 0.5)} \\
\midrule
Gemini~1.5~Pro & 0.335 & (0.127) & [0.086, 0.584] & 0.725 & $-0.002$ \\
Gemini~2.5~Pro & 0.330 & (0.134) & [0.068, 0.592] & 0.716 & $+0.001$ \\
Gemini~3.1~Pro & 0.305 & (0.122) & [0.066, 0.543] & 0.726 & $-0.001$ \\
GLM~5.2        & 2.134 & (0.691) & [0.781, 3.488] & 0.699 & $+0.010$ \\
GPT-4o         & $-0.077$ & (0.140) & [$-0.352$, 0.199] & 0.718 & $-0.002$ \\
GPT-5.1        & 0.288 & (0.095) & [0.101, 0.475] & 0.711 & $+0.001$ \\
GPT-5.4        & 0.238 & (0.139) & [$-0.034$, 0.510] & 0.729 & $-0.001$ \\
GPT-5.5        & 0.516 & (0.148) & [0.225, 0.807] & 0.717 & $-0.001$ \\
GPT-5.6-Luna   & 0.732 & (0.196) & [0.347, 1.117] & 0.718 & $+0.001$ \\
\bottomrule
\end{tabularx}
\begin{tablenotes}[flushleft]\footnotesize
\item \textit{Note:} Reported AI use is regressed on the direct (E1) and latent (E2$+$E3) components, each entered as a percentile rank; the implied weight is the ratio of their coefficients, with a delta-method standard error. The components correlate at 0.53--0.94, making the ratio unstable; GPT-4o's negative value reflects its denominator rather than a protective latent component. The last columns compare out-of-sample AUC under the implied weight and $\omega=0.5$; no gap exceeds 0.01. $N=5{,}772$; survey weighted, with standard errors clustered on 4-digit occupation.
\item \textit{Source:} SES~2023--24, ages 20--65. Authors' calculations.
\end{tablenotes}
\end{threeparttable}
\end{table}

\FloatBarrier

The share of working time the task battery observes is estimated by regressing usual working hours on task load in SES~2017 and 2023--24, giving $\hat{S}$, the share of hours associated with the 44 tasks (\cref{tab:appB:coverage_soc1}). Coverage is lowest in occupations whose core activities rely on physical manipulation in non-digital settings. \Cref{tab:appB:coverage_bounds} reports two uniform scenarios for the unobserved remainder; both reduce discrimination relative to the index as constructed.

\begin{table}[htbp]
\centering
\begin{threeparttable}
\caption{\TabSevenCaption}
\label{tab:appB:coverage_soc1}
\setlength{\tabcolsep}{6pt}
\begin{tabularx}{\textwidth}{l *{3}{Y}}
\toprule
\textbf{SOC 2010 (1-digit)} & \textbf{Mean $\hat{S}$} \\
\midrule
1 Managers, Directors \& Senior Officials    & 0.405 \\
2 Professionals   & 0.364 \\
3 Associate Professionals \& Technicians   & 0.322 \\
4 Administrative \& Secretarial     & 0.293 \\
5 Skilled Trades     & 0.281 \\
6 Caring, Leisure \& Other Service    & 0.319 \\
7 Sales \& Customer Services    & 0.464 \\
8 Process, Plant \& Machine Operatives  & 0.174 \\
9 Elementary     & 0.250 \\
\midrule
\textbf{Total} & \textbf{0.327} \\
\bottomrule
\end{tabularx}
\begin{tablenotes}[flushleft]\footnotesize
\item \textit{Note:} Estimated share of working hours associated with the SES task battery, by 1-digit SOC~2010 group. Survey and non-response weights applied.
\item \textit{Source:} SES~2017 and 2023--24 (GB). Authors' calculations.
\end{tablenotes}
\end{threeparttable}
\end{table}

\FloatBarrier

\begin{table}[htbp]
\centering
\begin{threeparttable}
\caption{Sensitivity to Working Time the Task Battery Does Not Observe}
\label{tab:appB:coverage_bounds}
\setlength{\tabcolsep}{6pt}
\begin{tabularx}{\textwidth}{l *{5}{>{\centering\arraybackslash}X}}
\toprule
 & \textbf{Index as} & \multicolumn{2}{c}{\textbf{Unobserved time}} & \multicolumn{2}{c}{\textbf{Unobserved time}} \\
 & \textbf{constructed} & \multicolumn{2}{c}{\textbf{AI-resilient}} & \multicolumn{2}{c}{\textbf{fully exposed}} \\
\cmidrule(lr){3-4} \cmidrule(lr){5-6}
\textbf{Rater} & \textbf{AUC} & \textbf{AUC} & \textbf{$p$} & \textbf{AUC} & \textbf{$p$} \\
\midrule
Gemini~1.5~Pro & 0.725 & 0.662 & $<$0.001 & 0.534 & $<$0.001 \\
Gemini~2.5~Pro & 0.713 & 0.667 & 0.008    & 0.544 & $<$0.001 \\
Gemini~3.1~Pro & 0.726 & 0.684 & 0.003    & 0.524 & $<$0.001 \\
GLM~5.2        & 0.701 & 0.660 & 0.014    & 0.524 & $<$0.001 \\
GPT-4o         & 0.719 & 0.647 & $<$0.001 & 0.504 & $<$0.001 \\
GPT-5.1        & 0.714 & 0.635 & $<$0.001 & 0.511 & $<$0.001 \\
GPT-5.4        & 0.729 & 0.647 & $<$0.001 & 0.513 & $<$0.001 \\
GPT-5.5        & 0.715 & 0.648 & $<$0.001 & 0.518 & $<$0.001 \\
GPT-5.6-Luna   & 0.719 & 0.660 & $<$0.001 & 0.516 & $<$0.001 \\
\midrule
\textbf{Range} & 0.701--0.729 & 0.635--0.684 & & 0.504--0.544 & \\
\bottomrule
\end{tabularx}
\begin{tablenotes}[flushleft]\footnotesize
\item \textit{Note:} AUC measures how well each index orders workers by reported AI-software use; 0.5 is chance. The constructed index normalises over the 44 observed activities. The two scenarios instead assign all unobserved time either no exposure, $G\hat{S}$, or full exposure, $G\hat{S}+(1-\hat{S})$, where $\hat{S}$ is the worker's observed-time share. Because $\hat{S}$ varies across workers, both scenarios reorder them. Each $p$ compares the scenario with the same rater's constructed index in a single \texttt{somersd} fit. $\hat{S}$ is re-estimated by 1-digit SOC group (weighted mean 0.304). $N=5{,}758$; survey weighted, with standard errors clustered on 4-digit occupation.
\item \textit{Source:} SES~2023--24, ages 20--65. Authors' calculations.
\end{tablenotes}
\end{threeparttable}
\end{table}

\FloatBarrier

\begin{table}[htbp]
\centering
\begin{threeparttable}
\caption{Exposure and Reported AI Software Use, by Rater}
\label{tab:appB:digitaltools_raters}
\setlength{\tabcolsep}{6pt}

\begin{tabularx}{\textwidth}{l *{3}{>{\centering\arraybackslash}X}}
\toprule
& \textbf{Exposure} & \textbf{+ other digital} & \textbf{+ other exposure} \\
\textbf{Rater} & \textbf{alone} & \textbf{tools} & \textbf{measures} \\
\midrule
Gemini~1.5~Pro & 0.243 \: (0.028) & 0.154 \: (0.026) & 0.183 \: (0.026) \\
Gemini~2.5~Pro & 0.226 \: (0.027) & 0.148 \: (0.024) & 0.159 \: (0.024) \\
Gemini~3.1~Pro & 0.250 \: (0.026) & 0.158 \: (0.023) & 0.186 \: (0.025) \\
GLM~5.2        & 0.186 \: (0.023) & 0.128 \: (0.020) & 0.107 \: (0.024) \\
GPT-4o         & 0.228 \: (0.030) & 0.153 \: (0.026) & 0.157 \: (0.025) \\
GPT-5.1        & 0.215 \: (0.026) & 0.143 \: (0.022) & 0.133 \: (0.024) \\
GPT-5.4        & 0.240 \: (0.026) & 0.161 \: (0.024) & 0.166 \: (0.023) \\
GPT-5.5        & 0.210 \: (0.025) & 0.139 \: (0.021) & 0.143 \: (0.027) \\
GPT-5.6-Luna   & 0.231 \: (0.028) & 0.158 \: (0.024) & 0.152 \: (0.023) \\
\midrule
Range          & 0.186--0.250 & 0.128--0.161 & 0.107--0.186 \\
Pooled         & 0.226 & 0.149 & 0.154 \\
\bottomrule
\end{tabularx}

\begin{tablenotes}[flushleft]\footnotesize
\item \textit{Note:} Rater-specific versions of \cref{tab:digitaltools,tab:ame_ai_adoption}. Entries are average marginal effects on reported AI-software use per interquartile range of exposure. Column~(1) enters exposure alone; column~(2) adds the worker's other three digital tools; column~(3) instead adds nine occupational exposure measures. All specifications control for task load, survey year, 1-digit occupation and demographics. The final row reports the pooled estimates from the body.
\item Standard errors in parentheses capture sampling variation within a rater and are clustered on 4-digit occupation; cross-rater sensitivity is shown by the range. All estimates have $p<0.01$. $N=5{,}683$.
\item \textit{Source:} SES~2023--24, ages 20--65. Authors' calculations.
\end{tablenotes}

\end{threeparttable}
\end{table}

\begin{table}[htbp]
\centering
\begin{threeparttable}
\caption{Rank Correlation of Exposure with Existing Occupational Measures}
\label{tab:appB:ses_gaisi_corr}
\setlength{\tabcolsep}{6pt}

\begin{tabularx}{\textwidth}{l *{2}{>{\centering\arraybackslash}X}}
\toprule
\textbf{Measure} & \textbf{Mean $\rho$} & \textbf{Range across raters} \\
\midrule
\multicolumn{3}{l}{\textit{Pre-generative-AI constructs}} \\
Automation probability \parencite{Frey2017TheComputerisation} & $-0.580$ & $[-0.621,\ -0.496]$ \\
Robot exposure \parencite{Webb2020TheMarket}                  & $-0.472$ & $[-0.506,\ -0.422]$ \\
Software exposure \parencite{Webb2020TheMarket}               & $-0.005$ & $[-0.056,\ \ \ 0.049]$ \\
Suitability for machine learning \parencite{Brynjolfsson2018WhatEconomy} & 0.206 & $[\ \ 0.182,\ \ \ 0.233]$ \\
AI exposure \parencite{Webb2020TheMarket}                     & 0.310 & $[\ \ 0.267,\ \ \ 0.335]$ \\
\addlinespace
\multicolumn{3}{l}{\textit{Generative-AI constructs, human-derived}} \\
Occupational AI exposure \parencite{Felten2021OccupationalUses} & 0.801 & $[\ \ 0.758,\ \ \ 0.833]$ \\
Generative AI exposure, human raters \parencite{Eloundou2024GPTsLLMs} & 0.758 & $[\ \ 0.737,\ \ \ 0.782]$ \\
Occupational LLM exposure \parencite{Felten2023OccupationalAI} & 0.750 & $[\ \ 0.701,\ \ \ 0.791]$ \\
\addlinespace
\multicolumn{3}{l}{\textit{Generative-AI construct, LLM-generated}} \\
Generative AI exposure, GPT-4 raters \parencite{Eloundou2024GPTsLLMs} & 0.753 & $[\ \ 0.702,\ \ \ 0.784]$ \\
\bottomrule
\end{tabularx}

\begin{tablenotes}[flushleft]\footnotesize
\item \textit{Note:} Spearman correlations between worker-level GAISI and established occupation-level exposure measures, computed under each rater. Columns report the mean and range of the nine correlations. Measures are grouped by construct and matched to workers through 4-digit SOC~2010, SOC~2020 or ISCO-08. With about 5,770 workers, all except software exposure differ from zero at $p<0.001$; tests are omitted because they add no separation.
\item \textit{Source:} SES~2023--24, ages 20--65. Authors' calculations.
\end{tablenotes}

\end{threeparttable}
\end{table}

\begin{table}[htbp]
\centering
\begin{threeparttable}
\caption{Porting the UK Exposure Ranking to PIAAC Workers}
\label{tab:appB:piaac_port}
\setlength{\tabcolsep}{6pt}
\begin{tabularx}{\textwidth}{l *{4}{Y}}
\toprule
 & \textbf{UK ranking} & \textbf{$R^2$} & \textbf{Partial $R^2$} & \textbf{Share of (3)} \\
 & (1) & (2) & (3) & (4) \\
\midrule
\multicolumn{5}{l}{\textit{Panel A. What the ported ranking adds}} \\
Controls only                  &               & 0.275 &       &       \\
\quad + UK exposure ranking    & 0.728 (0.063) & 0.643 & 0.507 & 0.727 \\
\quad + occupation $\times$ industry dummies &  & 0.780 & 0.697 & 1.000 \\
\addlinespace
No controls                    & 0.797 (0.064) & 0.628 &       &       \\
Occupation only, no industry   & 0.698 (0.169) & 0.589 & 0.433 & 0.634 \\
\midrule
\multicolumn{5}{l}{\textit{Panel B. By rater}} \\
Gemini~1.5~Pro   & 0.741 (0.066) & & 0.504 & \\
Gemini~2.5~Pro   & 0.706 (0.059) & & 0.501 & \\
Gemini~3.1~Pro   & 0.734 (0.051) & & 0.531 & \\
GLM~5.2          & 0.739 (0.064) & & 0.515 & \\
GPT-4o           & 0.736 (0.063) & & 0.500 & \\
GPT-5.1          & 0.748 (0.055) & & 0.540 & \\
GPT-5.4          & 0.731 (0.053) & & 0.527 & \\
GPT-5.5          & 0.736 (0.054) & & 0.523 & \\
GPT-5.6-Luna     & 0.740 (0.063) & & 0.504 & \\
\midrule
\textbf{Ensemble average} & \textbf{0.728 (0.063)} & & \textbf{0.507} & \\
\bottomrule
\end{tabularx}
\begin{tablenotes}[flushleft]\footnotesize
\item \textit{Note:} UK GAISI is averaged within 1-digit occupation $\times$ broad-industry cells in the English SES, attached to PIAAC workers in the same cells, and percentile-ranked. The outcome is each worker's percentile-ranked PIAAC exposure, so coefficients compare interquartile ranges. Controls are country, sex, education, age, literacy and numeracy. Partial $R^2$ is the share of residual variation explained by the added term; column~(4) scales it to the partial $R^2$ from a full set of cell dummies. Panel~B repeats the controlled specification by rater.
\item Standard errors in parentheses are clustered on 35 occupation $\times$ industry cells. Wild-cluster-bootstrap $p$-values are $<0.001$ for the ensemble and 0.008 for the occupation-only specification (9 clusters). $N=84{,}560$ workers in 22 countries.
\item The UK ranking takes 35 cell-level values, so the results describe agreement between cells rather than worker-level predictive power.
\item \textit{Source:} OECD Survey of Adult Skills (PIAAC) Cycle~2 and SES~2023--24. Authors' calculations.
\end{tablenotes}
\end{threeparttable}
\end{table}

\FloatBarrier

\begin{table}[htbp]
\centering
\begin{threeparttable}
\caption{Demographic Association with Residual Exposure, by Rater}
\label{tab:appB:gaisi_bias}
\setlength{\tabcolsep}{6pt}
\begin{tabularx}{\textwidth}{l *{4}{Y}}
\toprule
\textbf{Rater} & \textbf{$F$} & \textbf{$p$} & \textbf{$R^2$} & \textbf{Largest $|\beta|$} \\
\midrule
Gemini~1.5~Pro   & 3.44 & 0.009 & 0.007 & 0.011 \\
Gemini~2.5~Pro   & 2.00 & 0.094 & 0.004 & 0.012 \\
Gemini~3.1~Pro   & 5.47 & 0.000 & 0.010 & 0.012 \\
GLM~5.2          & 4.86 & 0.001 & 0.009 & 0.014 \\
GPT-4o           & 3.54 & 0.008 & 0.006 & 0.015 \\
GPT-5.1          & 1.90 & 0.111 & 0.004 & 0.010 \\
GPT-5.4          & 4.79 & 0.001 & 0.008 & 0.009 \\
GPT-5.5          & 3.86 & 0.004 & 0.008 & 0.007 \\
GPT-5.6-Luna     & 4.54 & 0.001 & 0.007 & 0.008 \\
\midrule
\textbf{Ensemble average} & \textbf{4.21} & \textbf{0.002} & \textbf{0.008} & \textbf{0.009} \\
\bottomrule
\end{tabularx}
\begin{tablenotes}[flushleft]\footnotesize
\item \textit{Note:} Exposure rank is first residualised on task profile, managerial status, task load and 2-digit occupation, then regressed on sex, age in decades, non-white ethnicity and tertiary education. The table reports their joint test, $R^2$, and largest absolute coefficient in interquartile-range units. The final row uses mean exposure across raters. Survey weighted; standard errors clustered on 4-digit occupation; $N=5{,}784$.
\item \textit{Source:} SES~2023--24, ages 20--65. Authors' calculations.
\end{tablenotes}
\end{threeparttable}
\end{table}

\section{Appendix: Labour-Market Estimates by Rater}
\label{app:C:application}

This appendix reports the two labour-market specifications of
\cref{ss:macro_signals} estimated separately under each of the nine LLM
raters. Both are on the percentile-rank scale, with coefficients reported
per interquartile range of the exposure ranking.

\begin{table}[htbp]\centering
      \begin{threeparttable}
      \caption{Pay and Generative AI Exposure across the Rater Panel}
      \label{tab:appC:ai_price_diffs}
      
      \begin{tabularx}{\textwidth}{l *{4}{>{\centering\arraybackslash}X}}
      \toprule
       & \multicolumn{3}{c}{Panel A: QLFS 2017--2026 Q1} & Panel B: SES 2017 vs 2023/24 \\
      \cmidrule(lr){2-4}\cmidrule(lr){5-5}
       & \multicolumn{3}{c}{\footnotesize vs post-pandemic 2022} & \\
      Rater & 2017--19 & 2020--21 & 2023--26 Q1 & Exposure \\
            &          &          &          & $\times$ 2023/24 \\
      \midrule
      Gemini 1.5 Pro & 0.029\sym{***} & 0.010 & $-$0.020\sym{**} & $-$0.051 \\
       & (0.007) & (0.006) & (0.007) & (0.027) \\
      \addlinespace
      Gemini 2.5 Pro & 0.023\sym{**} & 0.007 & $-$0.021\sym{**} & $-$0.056\sym{*} \\
       & (0.007) & (0.006) & (0.007) & (0.026) \\
      \addlinespace
      Gemini 3.1 Pro & 0.028\sym{***} & 0.010 & $-$0.020\sym{**} & $-$0.039 \\
       & (0.007) & (0.006) & (0.007) & (0.028) \\
      \addlinespace
      GLM 5.2 & 0.030\sym{***} & 0.014 & $-$0.020\sym{**} & $-$0.065\sym{**} \\
       & (0.007) & (0.007) & (0.007) & (0.023) \\
      \addlinespace
      GPT-4o & 0.026\sym{***} & 0.007 & $-$0.023\sym{**} & $-$0.043 \\
       & (0.007) & (0.006) & (0.007) & (0.027) \\
      \addlinespace
      GPT-5.1 & 0.029\sym{***} & 0.011 & $-$0.019\sym{**} & $-$0.059\sym{*} \\
       & (0.007) & (0.007) & (0.007) & (0.025) \\
      \addlinespace
      GPT-5.4 & 0.023\sym{***} & 0.004 & $-$0.021\sym{**} & $-$0.041 \\
       & (0.007) & (0.006) & (0.007) & (0.025) \\
      \addlinespace
      GPT-5.5 & 0.028\sym{***} & 0.011 & $-$0.018\sym{*} & $-$0.057\sym{*} \\
       & (0.007) & (0.007) & (0.007) & (0.023) \\
      \addlinespace
      GPT-5.6-Luna & 0.025\sym{***} & 0.007 & $-$0.022\sym{**} & $-$0.038 \\
       & (0.007) & (0.007) & (0.007) & (0.028) \\
      \midrule
      Rubin-pooled & 0.027\sym{***} & 0.009 & $-$0.020\sym{**} & $-$0.050 \\
       & (0.008) & (0.007) & (0.007) & (0.026) \\
      Rater share of variance & 0.139 & 0.207 & 0.047 & 0.141 \\
      Negative and $p<0.05$ & 0/9 & 0/9 & \textbf{9/9} & 4/9 \\
      Observations & \multicolumn{3}{c}{279{,}709} & 6{,}928 \\
      Occupation clusters & \multicolumn{3}{c}{89} & 89 \\
      \bottomrule
      \end{tabularx}
      
      \begin{tablenotes}[flushleft]
      \scriptsize
      \item \textit{Note:} Rater-specific and pooled coefficients in log points per interquartile range of exposure. \textbf{Panel A} interacts an occupation's mean rank with period indicators in QLFS log-pay regressions; 2022 is the reference. Controls are age and its square, sex, education and foreign-born status; 3-digit occupation is absorbed and standard errors clustered at that level. \textbf{Panel B} interacts worker-level exposure with the 2023/24 SES wave. Controls are sex interacted with age and its square, education, ethnicity, full-time, self-employment and task load; 3-digit occupation, region and 1-digit industry are interacted with wave. Survey weights are applied and standard errors clustered on 3-digit occupation. The panels use different exposure aggregation and should not be compared in magnitude. Pooled $p$-values are 0.0005, 0.213 and 0.005 in Panel A and 0.074 in Panel B. ``Rater share of variance'' is the between-rater share of the Rubin-combined variance. Standard errors in parentheses.
      \item \textit{Source:} QLFS 2017 to 2026~Q1; SES 2017 and 2023--24. Authors' calculations.
      \item \sym{*} $p<0.05$, \sym{**} $p<0.01$, \sym{***} $p<0.001$.
      \end{tablenotes}
      
      \end{threeparttable}
      \end{table}

\begin{table}[htbp]\centering
\begin{threeparttable}
\caption{Job Adverts and Generative AI Exposure across the Rater Panel}
\label{tab:appC:ai_demand_diffs}

\begin{tabularx}{\textwidth}{l *{5}{>{\centering\arraybackslash}X}}
\toprule
 & \multicolumn{2}{c}{Panel A: Stock of live adverts} & \multicolumn{3}{c}{Panel B: New postings} \\
\cmidrule(lr){2-3}\cmidrule(lr){4-6}
Rater & Pre & Post & Pre & Post & To \\
      & 2017--22 & 2023--25 & 2017--22 & 2023--25 & 2026 Q2 \\
\midrule
Gemini 1.5 Pro & 0.095\sym{***} & $-$0.172\sym{***} & 0.169\sym{***} & $-$0.169\sym{***} & $-$0.147\sym{***} \\
 & (0.010) & (0.009) & (0.009) & (0.009) & (0.009) \\
\addlinespace
Gemini 2.5 Pro & 0.080\sym{***} & $-$0.191\sym{***} & 0.148\sym{***} & $-$0.185\sym{***} & $-$0.176\sym{***} \\
 & (0.009) & (0.009) & (0.008) & (0.008) & (0.008) \\
\addlinespace
Gemini 3.1 Pro & 0.097\sym{***} & $-$0.183\sym{***} & 0.188\sym{***} & $-$0.162\sym{***} & $-$0.142\sym{***} \\
 & (0.010) & (0.009) & (0.009) & (0.009) & (0.009) \\
\addlinespace
GLM 5.2 & 0.099\sym{***} & $-$0.193\sym{***} & 0.198\sym{***} & $-$0.127\sym{***} & $-$0.105\sym{***} \\
 & (0.010) & (0.010) & (0.010) & (0.009) & (0.009) \\
\addlinespace
GPT-4o & 0.035\sym{***} & $-$0.187\sym{***} & 0.105\sym{***} & $-$0.185\sym{***} & $-$0.180\sym{***} \\
 & (0.010) & (0.010) & (0.009) & (0.009) & (0.009) \\
\addlinespace
GPT-5.1 & 0.093\sym{***} & $-$0.164\sym{***} & 0.169\sym{***} & $-$0.126\sym{***} & $-$0.106\sym{***} \\
 & (0.010) & (0.009) & (0.009) & (0.009) & (0.009) \\
\addlinespace
GPT-5.4 & 0.127\sym{***} & $-$0.158\sym{***} & 0.195\sym{***} & $-$0.129\sym{***} & $-$0.105\sym{***} \\
 & (0.009) & (0.008) & (0.008) & (0.008) & (0.008) \\
\addlinespace
GPT-5.5 & 0.092\sym{***} & $-$0.164\sym{***} & 0.173\sym{***} & $-$0.117\sym{***} & $-$0.095\sym{***} \\
 & (0.009) & (0.008) & (0.008) & (0.008) & (0.008) \\
\addlinespace
GPT-5.6-Luna & 0.141\sym{***} & $-$0.166\sym{***} & 0.222\sym{***} & $-$0.121\sym{***} & $-$0.094\sym{***} \\
 & (0.010) & (0.009) & (0.009) & (0.009) & (0.009) \\
\midrule
Pooled (Rubin) & 0.096\sym{*} & $-$0.175\sym{***} & 0.174\sym{***} & $-$0.147\sym{***} & $-$0.128\sym{**} \\
 & (0.033) & (0.017) & (0.036) & (0.031) & (0.037) \\
\addlinespace
Negative and $p<0.05$ & 0/9 & \textbf{9/9} & 0/9 & \textbf{9/9} & \textbf{9/9} \\
Implied \% per $+$30 points & & $-$10.0 & & $-$8.4 & $-$7.4 \\
Observations & \multicolumn{2}{c}{1{,}124{,}376} & \multicolumn{3}{c}{1{,}261{,}219} \\
\bottomrule
\end{tabularx}

\begin{tablenotes}[flushleft]
\footnotesize
\item \textit{Note:} Rater-specific estimates from regressions of log adverts plus one on exposure rank interacted with quarter. The panel is 3-digit SOC~2020 $\times$ local authority $\times$ quarter; 2022~Q3 is the reference and pre-2019 quarters are binned. ``Pre'' averages 2017~Q2--2022~Q2, ``Post'' averages 2023~Q1--2025~Q2, and the final column extends the flow series through 2026~Q2. Coefficients are per interquartile range; the implied change per 30 percentile points is $\exp(0.6\hat{\beta})-1$. Models absorb occupation $\times$ area, 1-digit occupation $\times$ quarter and region $\times$ quarter; standard errors are clustered on occupation $\times$ area. The pooled row uses the Rubin-style approach. Panel~A is the discontinued stock series and Panel~B new postings; June~2026 is excluded because 30\% of cells are suppressed.
\item \textit{Source:} ONS Labour demand volumes by SOC, releases 202507 (stock) and 202606 (flow); SES~2023--24. Authors' calculations.
\item \sym{*} $p<0.05$, \sym{**} $p<0.01$, \sym{***} $p<0.001$.
\end{tablenotes}

\end{threeparttable}
\end{table}

\begin{table}[htbp]\centering
\begin{threeparttable}
\caption{Telework-Adjusted New-Posting Gradients by Broad Period}
\label{tab:appC:vacancy_telework}
\begin{tabularx}{\textwidth}{l *{3}{>{\centering\arraybackslash}X}}
\toprule
 & \multicolumn{3}{c}{vs post-pandemic 2022} \\
\cmidrule(lr){2-4}
 & \textbf{Pre-pandemic} & \textbf{Pandemic} & \textbf{Post-GPT} \\
\textbf{Rater} & \textbf{2017--19} & \textbf{2020--21} & \textbf{2023--26 Q2} \\
\midrule
Gemini~1.5~Pro & 0.156\sym{***} & 0.077\sym{***} & 0.088\sym{***} \\
 & (0.014) & (0.010) & (0.010) \\
Gemini~2.5~Pro & 0.146\sym{***} & 0.045\sym{***} & $-$0.015 \\
 & (0.012) & (0.008) & (0.008) \\
Gemini~3.1~Pro & 0.231\sym{***} & 0.070\sym{***} & 0.108\sym{***} \\
 & (0.015) & (0.010) & (0.010) \\
GLM~5.2 & 0.204\sym{***} & 0.027\sym{**} & 0.135\sym{***} \\
 & (0.015) & (0.010) & (0.011) \\
GPT-4o & 0.016 & $-$0.015 & $-$0.008 \\
 & (0.013) & (0.009) & (0.009) \\
GPT-5.1 & 0.202\sym{***} & 0.036\sym{***} & 0.136\sym{***} \\
 & (0.014) & (0.009) & (0.009) \\
GPT-5.4 & 0.262\sym{***} & 0.089\sym{***} & 0.090\sym{***} \\
 & (0.012) & (0.008) & (0.009) \\
GPT-5.5 & 0.250\sym{***} & 0.060\sym{***} & 0.167\sym{***} \\
 & (0.014) & (0.009) & (0.009) \\
GPT-5.6-Luna & 0.313\sym{***} & 0.089\sym{***} & 0.143\sym{***} \\
 & (0.014) & (0.009) & (0.009) \\
\midrule
Pooled (Rubin-style) & 0.198 & 0.053 & 0.094 \\
 & (0.091) & (0.037) & (0.069) \\
Pooled $p$ & 0.061 & 0.180 & 0.210 \\
Rater share of variance & 0.98 & 0.94 & 0.98 \\
Positive and $p<0.05$ & 8/9 & 8/9 & 7/9 \\
\bottomrule
\end{tabularx}
\begin{tablenotes}[flushleft]\footnotesize
\item \textit{Note:} Exposure-rank interactions from the broad-period model of new postings, after APS teleworkability is given its own interaction in each displayed period. Post-pandemic 2022 is the reference. Coefficients are log points per interquartile range of exposure. Models absorb occupation $\times$ local area, 1-digit occupation $\times$ quarter and region $\times$ quarter; standard errors are clustered on occupation $\times$ area. The pooled row uses the Rubin-style approach, and ``rater share'' is the between-rater share of its combined variance. $N=1{,}252{,}931$ common cells.
\item \textit{Source:} ONS Labour demand volumes by SOC, release 202606; Annual Population Survey, July~2021--June~2022; SES~2023--24. Authors' calculations.
\item \sym{*} $p<0.05$, \sym{**} $p<0.01$, \sym{***} $p<0.001$.
\end{tablenotes}
\end{threeparttable}
\end{table}

\FloatBarrier


\end{appendices}

\end{document}